\documentclass[%
 reprint,
 superscriptaddress,
 amsmath,amssymb,
 aps,
 twocolumn,
prb,
floatfix,
]{revtex4-2}

\usepackage[english]{babel}
\usepackage[utf8]{inputenc}
\usepackage{amsthm}
\usepackage{mathtools}
\usepackage{physics}
\usepackage{xcolor}
\usepackage{graphicx} 
\usepackage[export]{adjustbox}
\usepackage{placeins}
\usepackage[T1]{fontenc}
\usepackage{csquotes}
\usepackage{bm}
\usepackage{multirow}
\usepackage{amsmath}
\usepackage{amsfonts}
\usepackage[ruled,vlined]{algorithm2e}
\usepackage{comment}
\usepackage{color}
\usepackage{tabularx}
\usepackage{adjustbox}
\usepackage{bbm}
\usepackage{float}
\usepackage{array}
\usepackage[caption=false]{subfig}
\usepackage[normalem]{ulem}
\usepackage{hyperref} 
\hypersetup{allcolors=blue, colorlinks=true}

\begin{document}

\title{Percolation renormalization group analysis of confinement in \texorpdfstring{$\mathbb{Z}_2$}{Z2} lattice gauge theories}

\author{Gesa D{\"u}nnweber}
\affiliation{Faculty of Physics, Arnold Sommerfeld Centre for Theoretical Physics (ASC),\\Ludwig-Maximilians-Universit{\"a}t M{\"u}nchen, Theresienstr.~37, 80333 M{\"u}nchen, Germany}
\affiliation{Munich Center for Quantum Science and Technology (MCQST), Schellingstr. 4, 80799 M{\"u}nchen, Germany}

\author{Simon M. Linsel}
\affiliation{Faculty of Physics, Arnold Sommerfeld Centre for Theoretical Physics (ASC),\\Ludwig-Maximilians-Universit{\"a}t M{\"u}nchen, Theresienstr.~37, 80333 M{\"u}nchen, Germany}
\affiliation{Munich Center for Quantum Science and Technology (MCQST), Schellingstr. 4, 80799 M{\"u}nchen, Germany}
\affiliation{Department of Physics, Harvard University, Cambridge MA 02138, USA}

\author{Annabelle Bohrdt}
\affiliation{Munich Center for Quantum Science and Technology (MCQST), Schellingstr. 4, 80799 M{\"u}nchen, Germany}
\affiliation{Institute of Theoretical Physics, University of Regensburg, Universit{\"a}tsstr.~31, 93053 Regensburg, Germany}

\author{Fabian Grusdt}
\affiliation{Faculty of Physics, Arnold Sommerfeld Centre for Theoretical Physics (ASC),\\Ludwig-Maximilians-Universit{\"a}t M{\"u}nchen, Theresienstr.~37, 80333 M{\"u}nchen, Germany}
\affiliation{Munich Center for Quantum Science and Technology (MCQST), Schellingstr. 4, 80799 M{\"u}nchen, Germany}

\date{\today}

\begin{abstract}

The analytical study of confinement in lattice gauge theories (LGTs) remains a difficult task to this day. Taking a geometric perspective on confinement, we develop a real-space renormalization group (RG) formalism for $\mathbb{Z}_2$ LGTs using percolation probability as a confinement order parameter. The RG flow we analyze is constituted by both the percolation probability and the coupling parameters.
We consider a classical $\mathbb{Z}_2$ LGT in two dimensions, with matter and thermal fluctuations, and analytically derive the confinement phase diagram. We find good agreement with numerical and exact benchmark results and confirm that a finite matter density enforces confinement at $T<\infty$ in the model we consider. Our RG scheme enables future analytical studies of $\mathbb{Z}_2$ LGTs with matter and quantum fluctuations and beyond.

\end{abstract}


\maketitle

\section{Introduction}

The confinement problem is of fundamental interest in a variety of theoretical models, most notably in quantum chromodynamics \cite{Greensite, Pasechnik2021}. Lattice gauge theories (LGTs), originally introduced to study quantum chromodynamics non-perturbatively \cite{Wilson}, are of vital importance in the study of confinement \cite{Kogut1983}.
A famous example is the Fradkin-Shenker model \cite{FradkinShenker}, which features a confined and a deconfined topological phase while hosting a local $\mathbb{Z}_2$ symmetry. It is especially intriguing because of its wide applicability \cite{Lammert1995, Kitaev2003} and its experimental accessibility in modern quantum simulation platforms \cite{Satzinger2021, Semeghini2021, Barbiero2019, Schweizer2019, Homeier2023}.

It remains challenging to define experimentally, numerically and analytically accessible order parameters to probe confinement. Well-known examples are Wilson loops \cite{Wilson}, Polyakov loops \cite{Polyakov1975}, t'Hooft loops \cite{tHooft1978} and the Fredenhagen-Marcu operator \cite{FredenhagenMarcu1983,FredenhagenMarcu1986}, of which only the latter is applicable in the presence of dynamical matter. However, even the Fredenhagen-Marcu order parameter suffers from severe numerical instability in certain regimes \cite{Xu2024, Linsel2024}.
Recently, a percolation-inspired order parameter has been proposed \cite{Linsel2024}, overcoming such instabilities. It further establishes a geometric perspective on confinement where $\mathbb{Z}_2$ electric strings percolate in the deconfined phase. Because of the scale invariance of percolation near phase transitions \cite{Reynolds1977,Essam1980}, it is possible to construct an effective renormalization group formalism to analytically obtain the phase diagram - which is the main goal of this paper.

The renormalization group (RG) \cite{Wilson1971, Kadanoff1966}, originally developed in the context of quantum field theories, has been widely successful in many areas of physics. It was famously used in condensed matter physics to derive scaling laws for the Kondo model \cite{Anderson1970}. RG was also applied to study the confinement problem in LGTs without dynamical matter \cite{Tomboulis2009, Braun2010, Ogilvie2011}.

\begin{figure}[t]
\centering
\leftline{a.}
{\includegraphics[width = 0.4\textwidth]{\empty 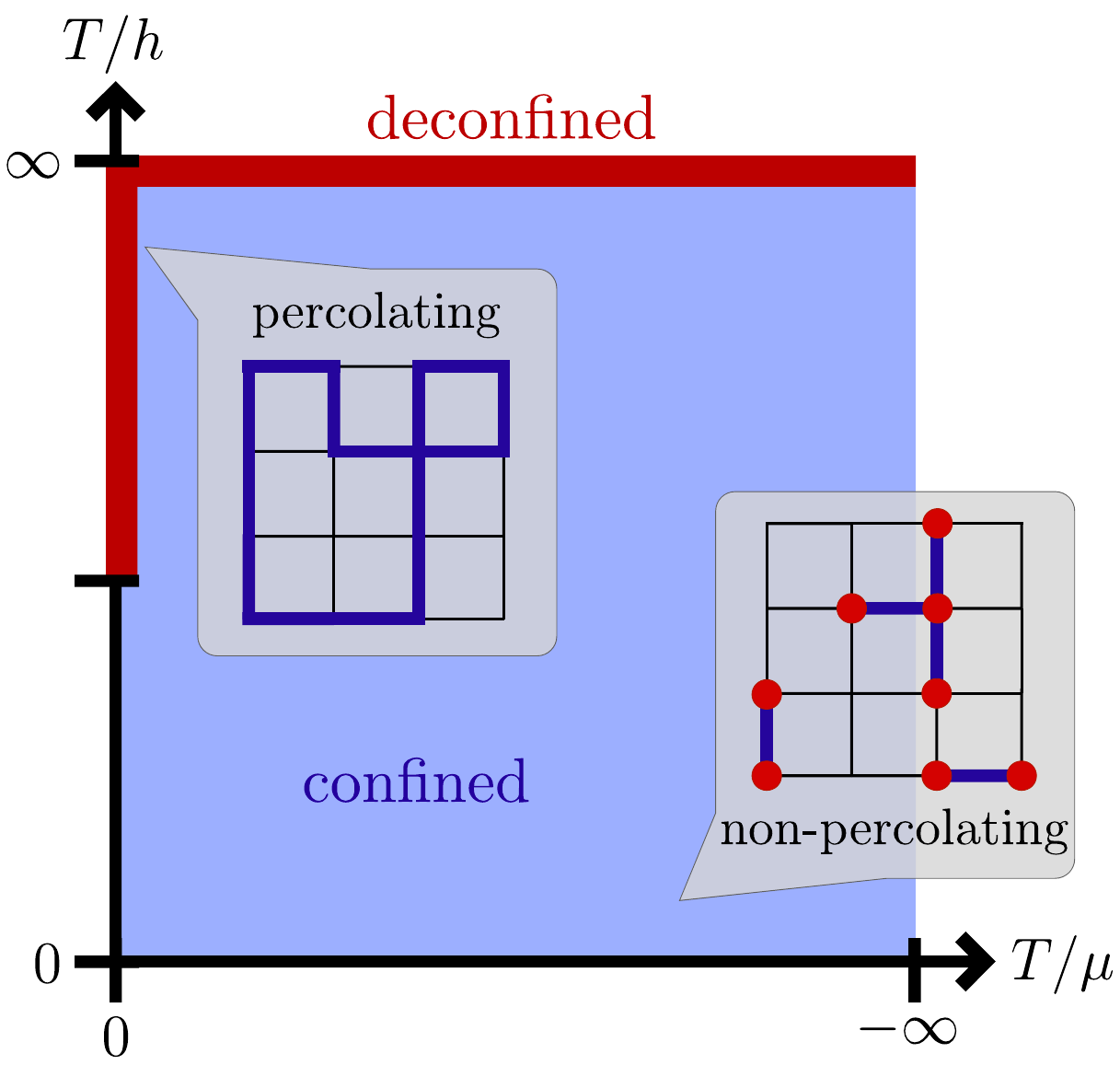}}
\leftline{b.}
{\includegraphics[width = 0.4\textwidth]{\empty 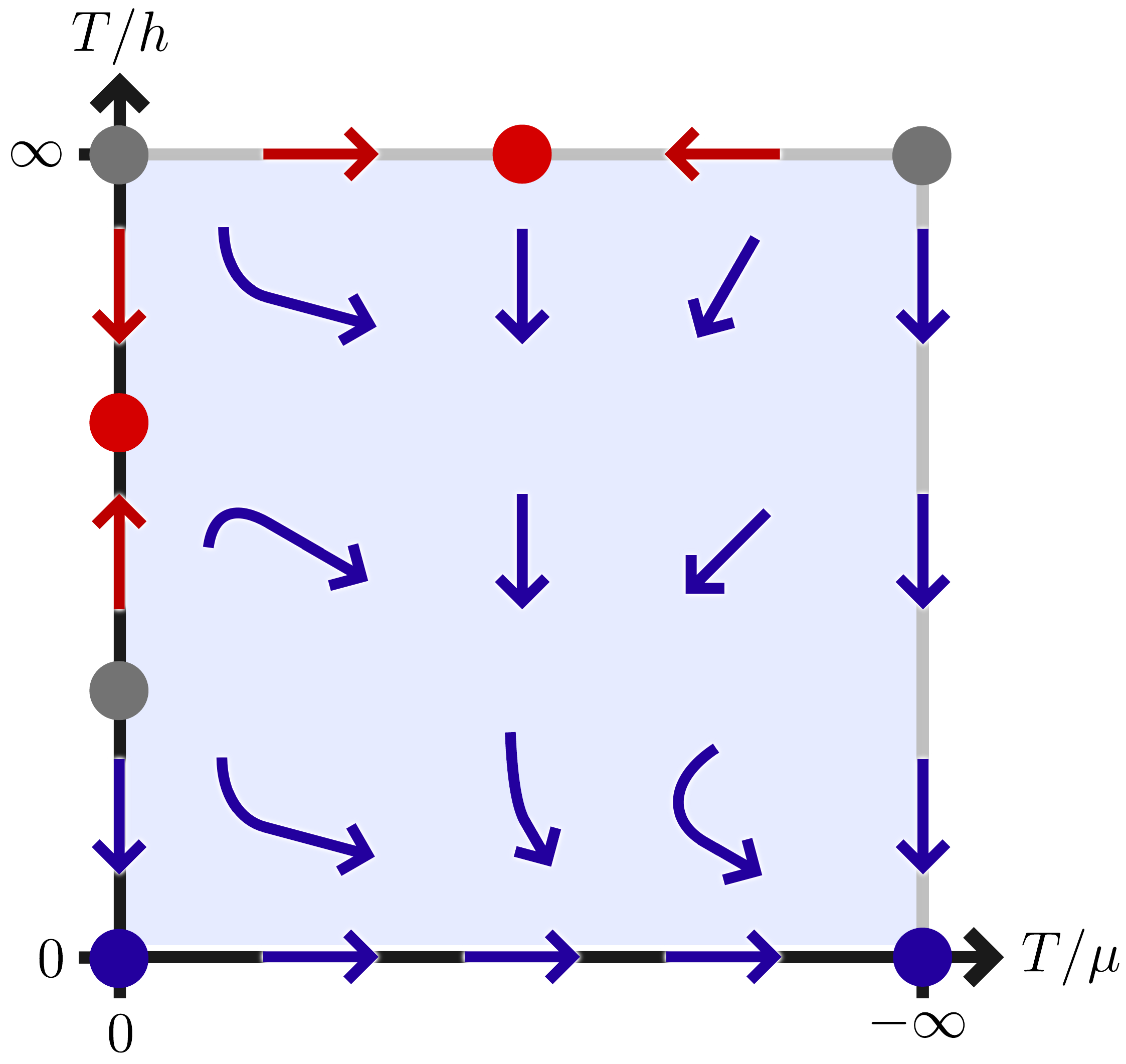}}
\caption{(a)~\label{fig:PhaseDiagram}\textbf{Confinement phase diagram} of the classical $\mathbb{Z}_2$ LGT we study.
We show that for positive $\beta h$ any nonzero particle density, i.e. any $T/\mu \neq 0$, leads to a deconfined system in which no percolating cluster of electric strings is formed. The system is deconfined in the limit $T / h = \infty$. The pure gauge system realized at $T/\mu = 0$ exhibits a confinement phase transition. (b)~\label{fig:RGflow_2D_wlims}\textbf{RG flow in the $\mu$-$h$ plane.} The phase diagram in the thermodynamic limit is constructed from this RG flow. \mbox{(Non-)}percolating fixed points are marked in red (blue); gray points correspond to unstable fixed points.
}
\end{figure}

In this paper we introduce a real-space RG formalism for $\mathbb{Z}_2$ lattice gauge theories with matter, featuring a simultaneous flow of coupling parameters and percolation probability. 
We apply the RG formalism to explain the confinement phase diagram of a classical $\mathbb{Z}_2$ LGT with fluctuating matter (see Fig.~\ref{fig:PhaseDiagram}a).
Our analytical results are in good agreement with earlier numerical studies of this model \cite{Linsel2024}. In particular we provide analytical understanding why any non-zero concentration of matter excitations at finite temperature leads to confinement in this model. Moreover, we go beyond earlier studies \cite{Linsel2024} by considering the regime of vanishing string tension $h = 0$ (or infinite temperature $T/h = \infty$) where we predict a thermally deconfined state with percolating strings, see Fig.~\ref{fig:PhaseDiagram}a. We confirm this prediction by numerically exact Monte Carlo simulations.

Our RG method lays the foundation to study percolation-based confinement in more complicated models. For example, the Fradkin-Shenker model \cite{FradkinShenker} could be analyzed, and the interplay of thermal and quantum fluctuations can be explored. This adds further analytical understanding of confinement of dynamical charges in $\mathbb{Z}_2$ LGTs and beyond: For instance, the critical exponents associated with the confinement-deconfinement transition that we find at unstable fixed-points in the RG flow diagram (see Fig.~\ref{fig:PhaseDiagram}b) can be analyzed.

\section{\texorpdfstring{$\mathbb{Z}_2$}{Z2} Lattice Gauge Theory}

We consider the classical two-dimensional $\mathbb{Z}_2$ lattice gauge theory in the electric field basis 
\begin{equation}
\hat{H} = - h \sum_{\langle\mathbf{i}, \mathbf{j}\rangle}\hat{\tau}^x_{\langle\mathbf{i}, \mathbf{j}\rangle} \end{equation} where $\hat{\tau}^x_{\langle\mathbf{i}, \mathbf{j}\rangle} = {\pm 1}$ is the electric field on the bond connecting neighboring sites $\mathbf{i}$ and $\mathbf{j}$ on the square lattice. Sites can be locally occupied by hard-core bosonic matter, subject to a $\mathbb{Z}_2$ Gauss's law. This defines a classical statistical mechanics problem with Boltzmann weight $ w = \frac{1}{Z}e^{-\beta H}$, temperature $T = 1/ \beta$ and partition function $Z = \sum e^{-\beta H}$, where the sum is over all gauge-invariant states.

\begin{figure}[t]
\centering
\leftline{a.}
{\includegraphics[width = 0.48 \textwidth]{\empty 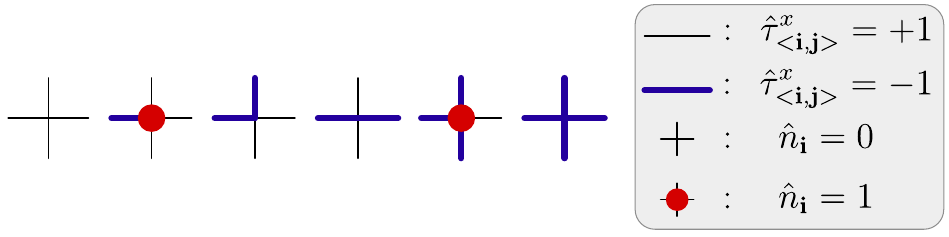}}
\leftline{b.}
{\includegraphics[width = 0.48 \textwidth]{\empty 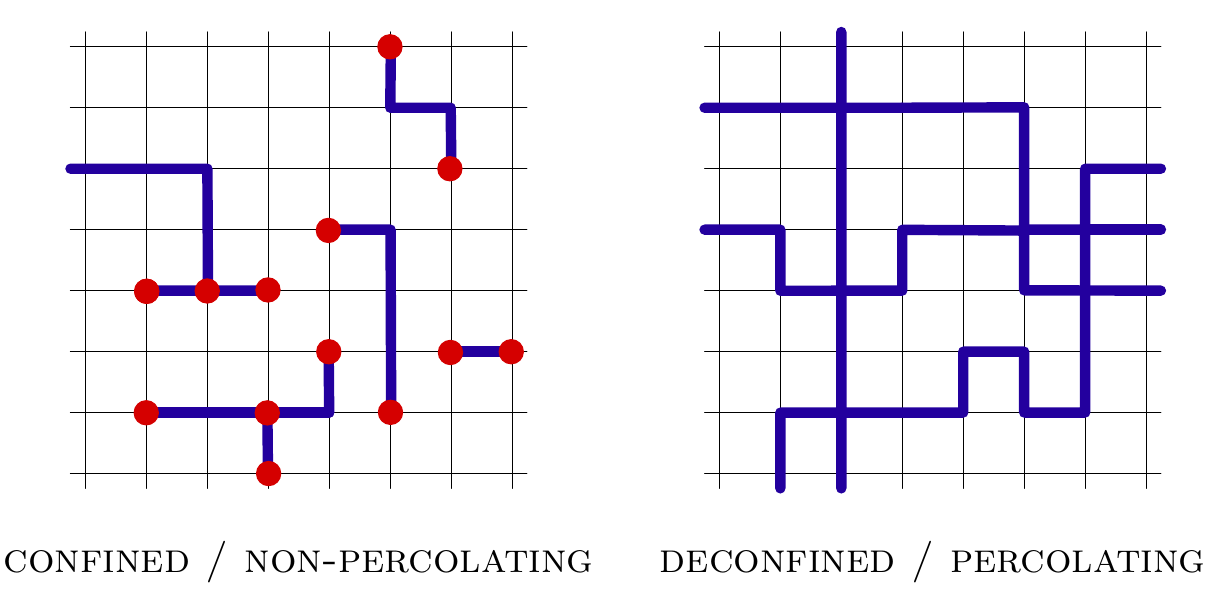}}
\caption{(a)\label{fig:GaussLaw}~\textbf{$\mathbb{Z}_2$ Gauss's law on the square lattice.} We illustrate the allowed configurations under Gauss's law on the square lattice in the electric field basis. (b)\label{fig:percolation}~\textbf{Percolation probability as an order parameter for confinement.} 
In the deconfined phase, there exists a percolating cluster of $\mathbb{Z}_2$ electric strings. In the confined regime, matter is bound in local $\mathbb{Z}_2$-neutral clusters (hadrons). We show that any nonzero charge density prohibits thermal deconfinement.
}
\end{figure}

The local $\mathbb{Z}_2$ symmetry generator is given by
\begin{equation}
    \hat{G}_\mathbf{j} = (-1)^{\hat{n}_\mathbf{j}} \prod_{\mathbf{i}: \langle\mathbf{i}, \mathbf{j}\rangle}\hat{\tau}^x_{\langle\mathbf{i}, \mathbf{j}\rangle}.
\end{equation} 
Choosing a gauge sector yields locally conserved quantities -- the charges $g_\mathbf{j}$. We choose the \textit{physical sector} with $g_\mathbf{j} = 1$ at all sites $\mathbf{j}$, which is interpreted as having no background charges. This local constraint is known as \textit{Gauss's law}. Thus, the hard-core matter particles with density $\hat{n}_{\mathbf{j}} = \hat{a}^{\dag}_{\mathbf{j}} \hat{a}_{\mathbf{j}}$ must have an odd number of adjoining electric strings $\hat{\tau}^x_{\langle\mathbf{i}, \mathbf{j}\rangle} = -1$. Fig.~\ref{fig:GaussLaw}a illustrates the $\mathbb{Z}_2$ Gauss's law on the square lattice.

To study this model analytically, we use its grand canonical variant
\begin{eqnarray}
\hat{H} &&= - h \sum_{\langle\mathbf{i}, \mathbf{j}\rangle}\hat{\tau}^x_{\langle\mathbf{i}, \mathbf{j}\rangle} - \mu \sum_{\mathbf{j}} \hat{n}_{\mathbf{j}} \nonumber\\
&&= - h \sum_{\langle\mathbf{i}, \mathbf{j}\rangle}\hat{\tau}^x_{\langle\mathbf{i}, \mathbf{j}\rangle} + \mu \sum_{\mathbf{j}} \prod_{\mathbf{i}: \langle\mathbf{i}, \mathbf{j}\rangle}\hat{\tau}^x_{\langle\mathbf{i}, \mathbf{j}\rangle} + \text{const.}, \label{eqn:gcHam}
\end{eqnarray} 
where $\mu$ is a chemical potential used to fix the matter density. By Gauss's law the latter can be expressed as $(-1)^{\hat{n}_{\mathbf{j}}} = \prod_{\mathbf{i}: \langle\mathbf{i}, \mathbf{j}\rangle}\hat{\tau}^x_{\langle\mathbf{i}, \mathbf{j}\rangle}$.
We refer to bonds with electric field $\hat{\tau}^x = -1$ as \emph{occupied}. These bonds form (electric) strings $\xi$ where $\hat{\tau}^x_l = -1$ $\forall l\in\xi$. We use the convention $h \geq 0$ and $\mu \leq 0$, i.e. the absence of electric strings and of matter particles is energetically preferred.

We define confinement geometrically in the electric field basis.
The percolation probability $\rho = \rho(\beta h, \beta \mu)$ is the probability that there exists a cluster of $\mathbb{Z}_2$ electric strings that extends across opposite ends of the lattice. In the \mbox{confined} regime, matter forms finite-size $\mathbb{Z}_2$ neutral clusters, resembling hadrons; in the \mbox{deconfined} regime, strings form a percolating cluster; see Fig.~\ref{fig:percolation}b. It has been demonstrated that $\rho$ is a suitable order parameter to probe confinement, including in cases with fluctuating matter \cite{Linsel2024}.

\section{Derivation of the confinement phase diagram}

\subsection{Renormalization group approach} \label{sec:RG}

We take a real-space RG approach where we perform a partial trace over the systems' degrees of freedom and assign each possible configuration a new \textit{block}-configuration on the renormalized system. Imposing that the total partition sum of the renormalized system is that of the original system,
\begin{eqnarray} Z &&= \sum_{\{\hat{\tau}^x_{\langle\mathbf{i}, \mathbf{j}\rangle}\}} e^{-\beta H(\{\hat{\tau}^x_{\langle\mathbf{i}, \mathbf{j}\rangle}\}, h, \mu)} \nonumber \\
&&\overset{!}{=} \sum_{\{\hat{\tau}^x_{\langle\mathbf{I}, \mathbf{J}\rangle}\}} e^{-\beta H'(\{\hat{\tau}^x_{\langle\mathbf{I}, \mathbf{J}\rangle}\}, h', \mu')}, \label{eqn:partsumrescaling}\end{eqnarray}
we obtain a new Hamiltonian $H'$ that must have (approximately) the same structure as the original Hamiltonian. The state $\{\hat{\tau}^x_{\langle\mathbf{i}, \mathbf{j}\rangle}\}$ before the RG map is called the micro-configuration and the state $\{\hat{\tau}^x_{\langle\mathbf{I}, \mathbf{J}\rangle}\}$ after renormalizing is called the block or macro-configuration. The RG procedure is repeated ad infinitum and the phase diagram in the thermodynamic limit can be constructed from the resulting parameter flow.

The challenge in constructing such a renormalization scheme for this $\mathbb{Z}_2$ LGT lies in the additional constraints that: (i) the block configuration must still satisfy Gauss's law and (ii) the percolation probability must be tractable throughout the renormalization.

\begin{figure}
\centering
\leftline{a.}
{\includegraphics[scale=0.5]{\empty 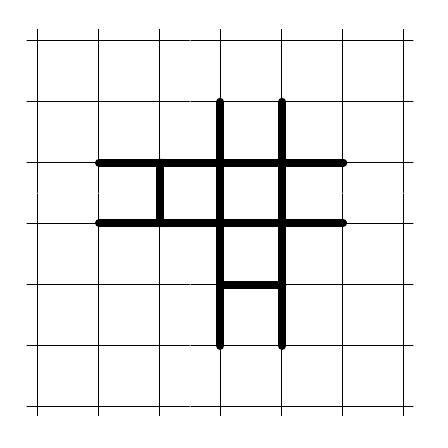}}
\leftline{b.}
{\includegraphics[scale=0.6]{\empty 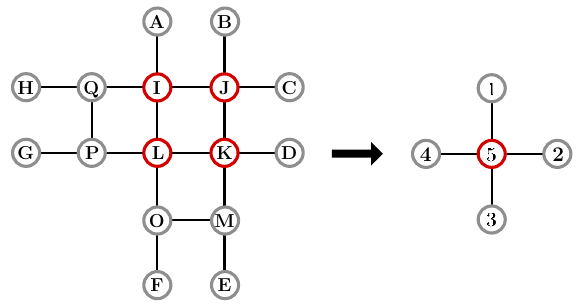}}
\leftline{c.}
{\includegraphics[scale=0.6]{\empty 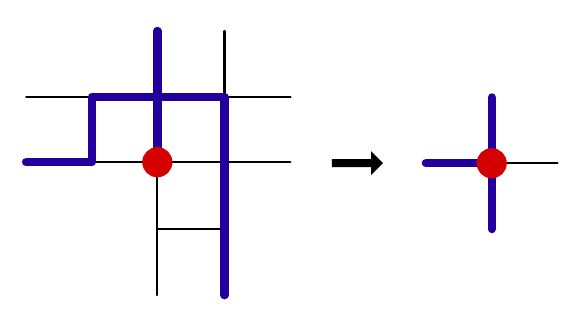}}
\leftline{d.}
{\includegraphics[scale=0.6]{\empty 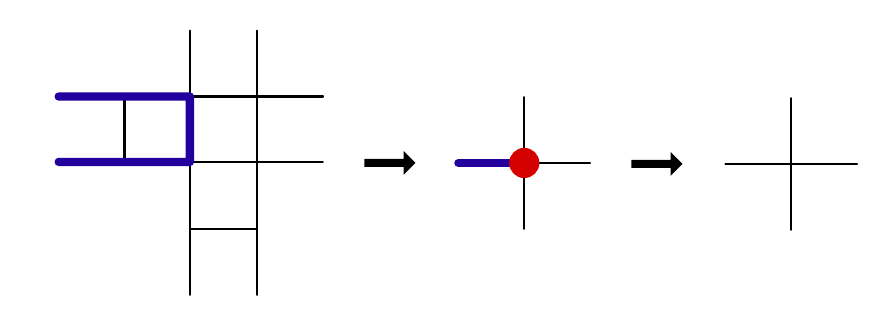}}
\caption{\label{fig:RGsketch} \textbf{Adjusted spanning cluster RG scheme.} The square lattice is divided into blocks (a) which are each assigned a renormalized/block configuration by checking for spanning clusters in each direction (b). We show the renormalization of an example configuration (c). In the case of a pure gauge configuration, the renormalized configuration is further adjusted to preserve particle number (d).
%
}
\end{figure}

Starting with requirement (ii), we mimic the \textit{spanning cluster rule} previously used \cite{Reynolds1977, Bernasconi1978} to analyze the Bernoulli percolation, which is the bond percolation on the square lattice.
Bernoulli percolation is realized in our model at $\mu = 0$, i.e. for independent links or without Gauss's law constraints. We then adapt this approach such that the density of matter particles obeying Gauss's law are approximately conserved, yielding a renormalized Hamiltonian that indeed takes the same form as the original Hamiltonian (to first order in $\beta$). This is the \textit{adjusted spanning cluster rule}:

\textbf{Step 1:} Divide the square lattice into blocks according to Fig.~\ref{fig:RGsketch}a. One such block is shown in Fig.~\ref{fig:RGsketch}b on the left. We use capital letters to label sites in the original configurations and numbers to label sites in the renormalized configuration.
\textbf{Step 2:} For each block, define the macro-configuration by setting $\hat{\tau}^x_{1,5} = -1$ if and only if there exists a path traversing only occupied bonds from $\{A,B\}$ to  $\{L,K\}$. If no such path exists, set $\hat{\tau}^x_{1,5} = +1$. Proceed similarly with paths from $\{C,D\}$ to $ \{I,L\}$ giving $\hat{\tau}^x_{2,5}$, from $\{F,E\}$ to $ \{L,K\}$ giving $ \hat{\tau}^x_{3,5}$ and from $\{G,H\}$ to $ \{I,L\}$ giving $ \hat{\tau}^x_{4,5}$. Hence, the block configuration has a horizontal electric string from vertex $2$ to vertex $5$ iff there exists a cluster spanning from the center to the right in the original configuration (\textit{spanning cluster rule}).
\textbf{Step 3:} Adjust the block configuration from step 2 to approximately preserve the particle density. Whenever the original configuration is pure gauge, i.e. whenever $n_\mathbf{j} = 0$ for the central sites $\mathbf{j} = I,J,K,L$, the block configuration is set to also have zero particles. 
To do this, note that the block configuration from step 2 corresponds to nonzero particle number precisely if it has an odd number of incoming electric strings (Gauss's law).
Thus there must be either one incoming string or three. In the former case, remove the string and in the latter case, add the fourth string. Either way, the block now has an even number of occupied bonds, and thus no particle at the central site.
\textbf{Step 4:} Introduce particles on the block configuration by following Gauss's law.

\subsection{Coupling parameter flow}

Now that the RG procedure is defined, it remains to track the flow of coupling parameters $h$, $\mu$ under this map.

The renormalized Hamiltonian $H'$ is determined via equation \ref{eqn:partsumrescaling} to be 
\begin{equation}
- \beta \hat{H}'\left(\left\{\hat{\tau}^x_{\langle\mathbf{I}, \mathbf{J}\rangle}\right\}\right) 
= \log(  
\sum_{\{\hat{\tau}^x_{\langle\mathbf{i}, \mathbf{j}\rangle}\}\text{*}} 
e^{-\beta \hat{H}\left(\left\{\hat{\tau}^x_{\langle\mathbf{i}, \mathbf{j}\rangle}\right\}, h, \mu\right)}  )
\label{eqn:newH}\end{equation} 
where the asterisk indicates that the sum is taken only over those micro-states $\{\hat{\tau}^x_{\langle\mathbf{i}, \mathbf{j}\rangle}\}$ for which the renormalization procedure yields the specific block-state $\{\hat{\tau}^x_{\langle\mathbf{I}, \mathbf{J}\rangle}\}$.

Imposing that $\hat{H}'$ takes a similar form to $\hat{H}$, specifically that interactions remain block-local, equation \ref{eqn:newH} holds not just for the entire lattice but also for states on a single block. For such a block the original state consists of $18$ bonds and the renormalized state of four bonds. The calculation can then be done exactly using symbolic computation to classify the $2^{18}$ micro-states according to their assigned macro-state and evaluating the respective sums.
For example,
\begin{equation}
\hspace{-3pt}\adjincludegraphics[valign=M,scale=0.26]{\empty 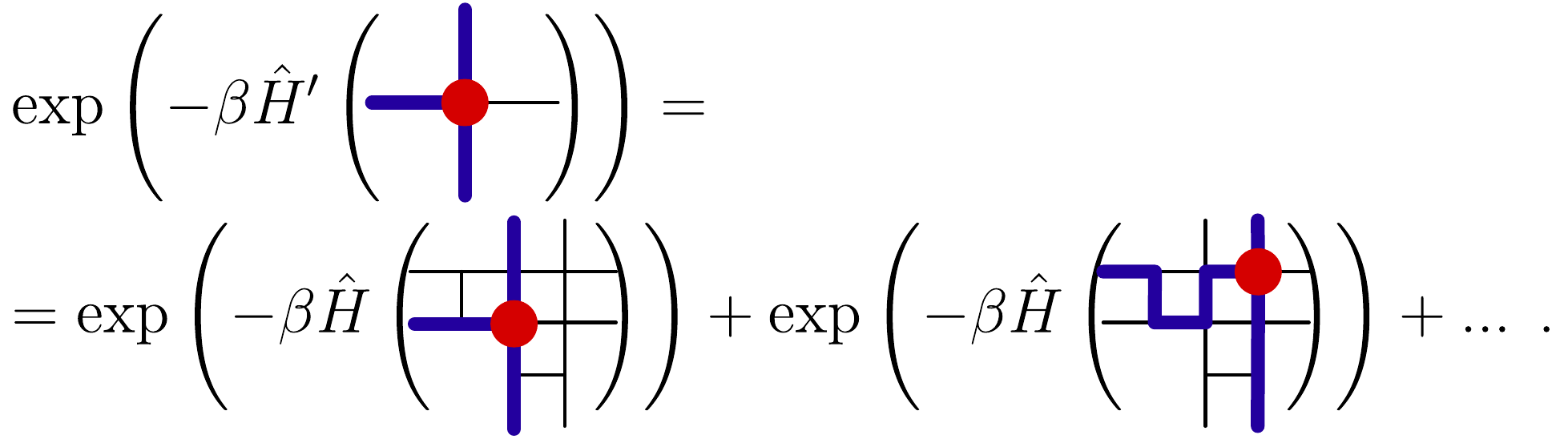}\hspace{3pt}
\end{equation}
We thus obtain formulas for the Boltzmann weights $\exp({- \beta \hat{H}'(\{\hat{\tau}^x_1, \hat{\tau}^x_2, \hat{\tau}^x_3, \hat{\tau}^x_4\})})$ of all possible block configurations shown in Fig.~\ref{fig:GaussLaw}. 
Keeping only the highest order terms in orders of $\beta$, the renormalized Hamiltonian
\begin{equation}
\hat{H}' = - h' \sum_{\langle\mathbf{i}, \mathbf{j}\rangle}\hat{\tau}^x_{\langle\mathbf{i}, \mathbf{j}\rangle} 
- \mu' \sum_{\mathbf{j}} \hat{n}_{\mathbf{j}} + \text{const.}\end{equation} 
again depends only on the coupling parameters $h',\mu'$
and the RG map $(h,\mu) \to (h',\mu')$ is obtained (see appendix \ref{sec:RG_appendix_flowderiv}). Fig.~\ref{fig:RGflow_2D} shows the resulting flow of the coupling parameters under the RG map.

\begin{figure}
\centering
{\includegraphics[width = 0.48\textwidth]{\empty 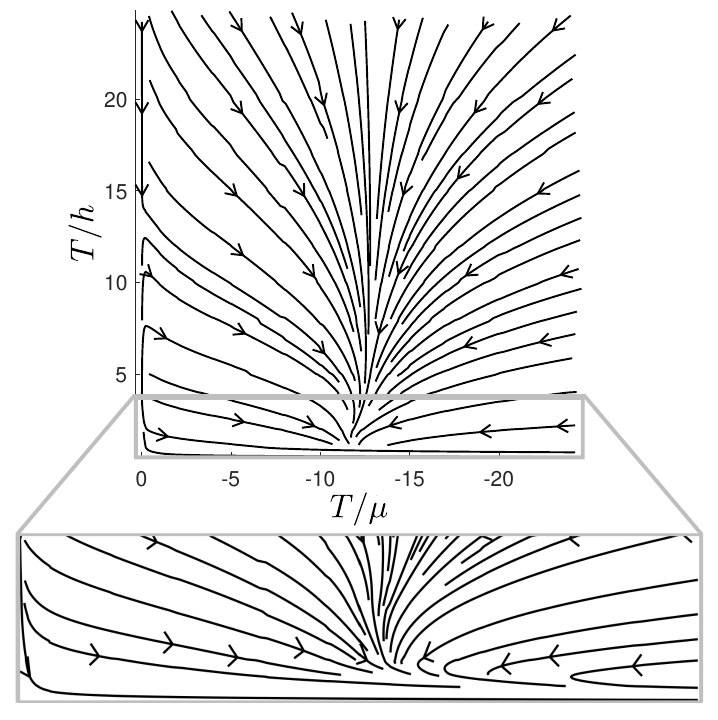}}
\caption{\label{fig:RGflow_2D} \textbf{RG flow, numerical plot.} All values of the coupling parameters $T/\mu \neq 0$, $T/h < \infty$ flow towards the non-percolating fixed point $T/\mu \to -\infty$, $T/h = 0$.}
\end{figure}

Fig.~\ref{fig:RGflow_2D_wlims}b depicts the fixed points of this flow including the limit cases corresponding to the axes of Fig.~\ref{fig:RGflow_2D}.
On the axis $T/h \to 0$, the only energetically allowed state is the state with no electric strings, so the system never percolates. (Alternatively, using the canonical Hamiltonian with fixed particle number or density, the limit $T/h \to 0$ yields a dimer state which also doesn't percolate.) 
The flow from points in the plane towards this limit indicates that the percolation probability is zero at all of these points.

In the limit $T/\mu = -\infty$, the $\hat{\tau}^x$ strings are independent, thus the system reduces to Bernoulli percolation. The percolation behavior is then determined by the bond occupation probability $p$. The critical probability on the square lattice is $p_c = 1/2$ \cite{Sykes1964,Kesten1980}. In the model at hand, $\displaystyle p = \frac{e^{-\beta h}}{e^{-\beta h} + e^{\beta h}} < \frac{1}{2}$ for $\beta h > 0$ (this corresponds to a two-level system realized by a single $\hat{\tau}^x$-link in a thermal ensemble). So the percolation probability is $0$, except at the point with $\beta h = 0$ (infinite temperature limit) where the percolation probability is $1$ and the system exhibits critical percolation. The critical probability depends crucially on the lattice geometry \cite{Sykes1964,Sykes1964exact} and number of spatial dimensions \cite{Wang2013}, so intuitively one should expect the percolation for finite $T / \mu$ to similarly  depends on these.

For the axis $T/h = \infty$, the system flows towards some fixed point $(T / \mu, T / h) = (T / \mu^*, \infty)$. In particular, the critically percolating point $(T / \mu, T / h) = (- \infty, \infty)$ flows towards the same fixed point, suggesting that
all $(T / \mu, T / h) = ( T / \mu, \infty)$ values yield a percolating system. As $\beta h$ is zero, there is no difference being made between strings and anti-strings in this limit, so any correlation effects stem from the preference of closed loops via $\beta \mu \neq 0$.

In the limit of a pure gauge theory with no particles, i.e. $T / \mu \to 0$, the system can be mapped to a classical Ising model on the dual lattice \cite{Fradkin1978, Wegner1971} with a percolation transition on this axis \cite{Linsel2024, Hastings2014}. The coupling parameter flow in Fig.~\ref{fig:RGflow_2D_wlims} exhibits several fixed points on this axis. 
We identify a stable, non-percolating (confined) fixed point at $T=0$; an unstable, i.e. critical, fixed point at a critical temperature $T_C = 3.48 h >0$ where the percolation transition of the pure gauge theory takes place; a stable, percolating (deconfined) fixed point above $T_C$; and an unstable, i.e. critical, fixed point at $T = \infty$ flowing towards the percolating (deconfined) state, demonstrating critical percolation at $T = \infty$.

\subsection{Percolation probability flow}

\begin{figure}
\centering
\leftline{a.}
{\includegraphics[scale=0.6]{\empty 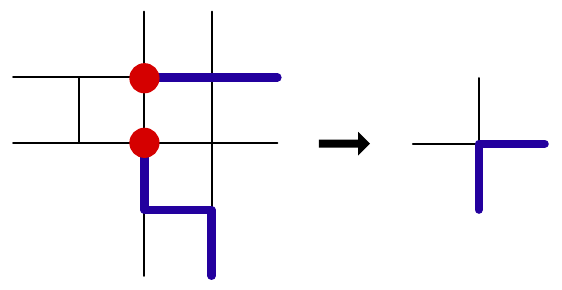}}
\leftline{b.}
{\includegraphics[scale=0.6]{\empty 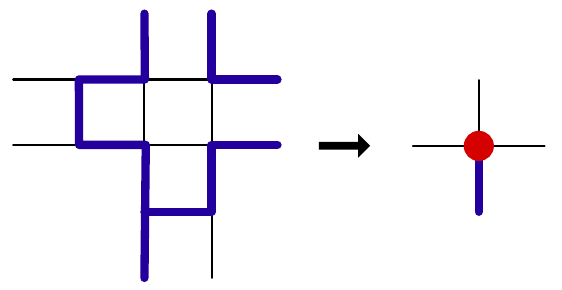}}
\caption{\label{fig:PercolationChange} \textbf{Percolation corrections.} 
We show that the percolation properties may be altered during an RG step. Non-percolating micro-states are sometimes mapped to percolating macro-states (a) or vice versa (b). The resulting percolation probability flow needs to be tracked alongside the flow of the coupling parameters to obtain the percolation phase diagram.
}
\end{figure}

So far we have only studied the RG flow of the renormalized coupling parameters $\beta \mu$ and $\beta h$ in the Hamiltonian. However, during each RG step, the percolation properties of the system may be altered by introducing or removing connections between electric string clusters. On the block level this corresponds to mapping a percolating micro-configuration to a non-percolating macro-configuration and vice versa. Both of these can happen with the adjusted spanning cluster rule we used (see Fig.~\ref{fig:PercolationChange}). Now, to extract the percolation transition, we revisit the implicit assumption made so far that the percolation probability is kept constant throughout the RG flow. In reality, further care must be taken to analyze how the renormalization procedure affects the percolation properties and we will see that this influences the conclusions drawn from the parameter flow.

We track the overall change $\delta \rho$ of the percolation probability $\rho$ which is equal to the average of the percolation change over all possible configurations weighted by the respective Boltzmann factors:
\begin{eqnarray}
\delta \rho (\beta h, \beta \mu) &&= \rho (\beta h, \beta \mu) - \rho (\beta h', \beta \mu') \\
&&= \frac{1}{Z(\beta h, \beta \mu)} \sum_{\text{states } \Phi} e^{-\beta \hat{H}(\Phi)} \cdot \left( \rho(\Phi) - \rho(\Phi')\right) . \nonumber
\end{eqnarray}
It is thus possible that during the flow from some point $(\beta h, \beta \mu)$ in coupling parameter space to a fixed point, the renormalization procedure has changed the clusters in such a way that a (non-)percolating configuration at the fixed point does not imply a (non-)percolating configuration at the starting point. Hence it is necessary to construct the flow of percolation probability alongside the RG flow of the coupling parameters. 

\begin{figure}
\centering
\includegraphics[width = 0.48 \textwidth]{\empty 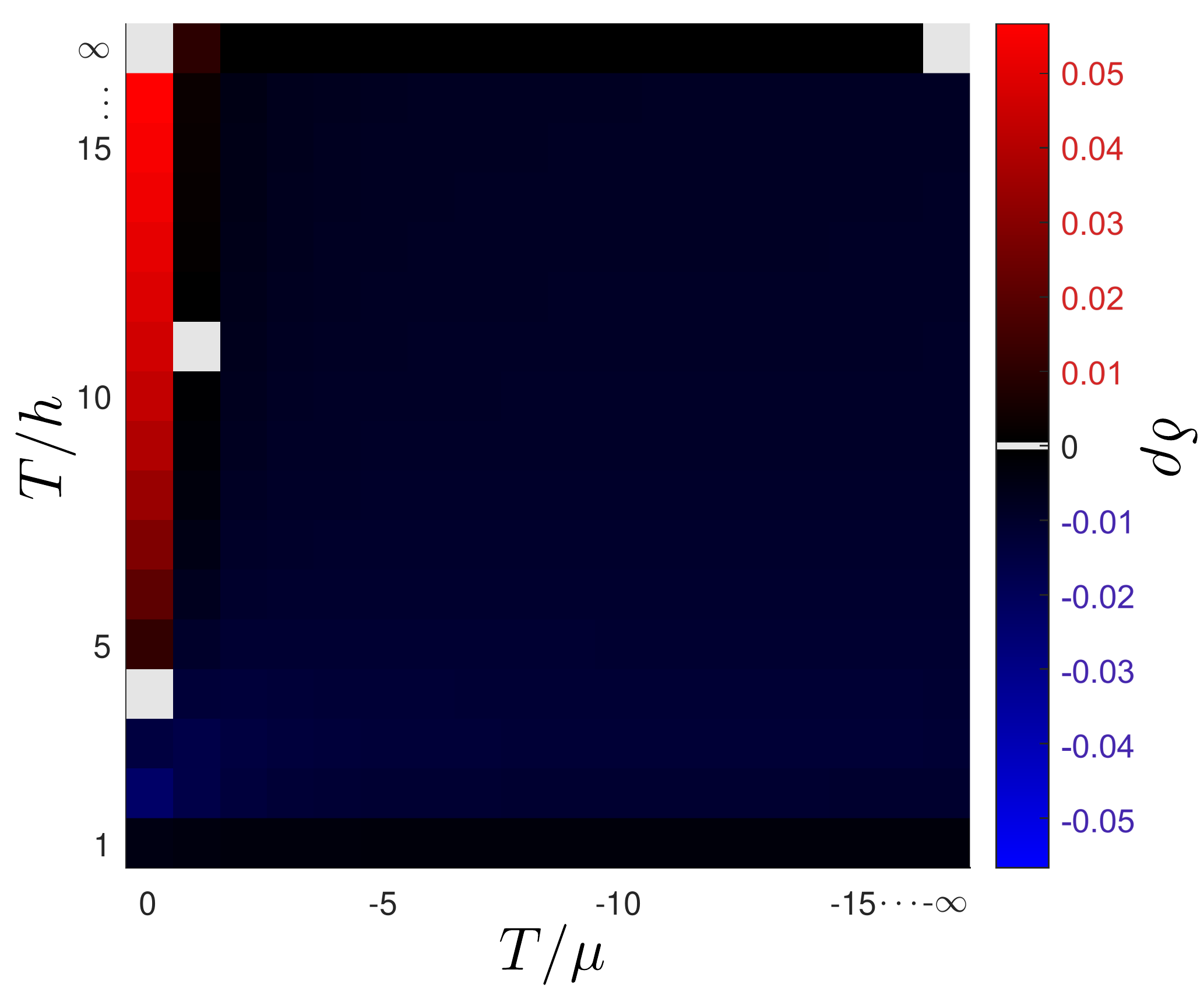}
\caption{\label{fig:PercCorrections} \textbf{Numerical values of the percolation corrections.} 
The percolation corrections are negative for $T/\mu \neq 0$, $T/h < \infty$. The percolation transition for the pure gauge theory with $T/\mu = 0$ is emphasized by the sign change of corrections on this axis. Overall, the percolation corrections are relatively small ($\delta \rho < 0.08$ $\forall (\beta h, \beta \mu)$), implying that the renormalization scheme can indeed be used to analyze percolation.
}
\end{figure}

To obtain the total change in percolation probability during the RG flow to some fixed point with known percolation probability, the changes from each of the RG steps are summed up, i.e.
\begin{eqnarray}
\Delta \rho (\beta h, \beta \mu) 
&&= \rho (\beta h, \beta \mu) - \rho_\infty  \nonumber\\
&&= \sum_{n=0}^\infty (\rho_{n} -\rho_{n+1} ) \nonumber\\
&&= \sum_{n= 0}^\infty \delta\rho_n .
\end{eqnarray}
Fig.~\ref{fig:PercCorrections} shows the corrections to the percolation obtained during the RG flow for various $\beta h, \beta \mu$. 

We analyze whether these percolation corrections correspond to a genuine change of percolation probability or if they can be neglected:
Consider the sign of the percolation corrections in various regions of the confinement phase diagram. If we flow towards a percolating fixed point, then positive percolation corrections mean that the initial system is more percolating than the renormalized system. In the case that percolation probability of the fixed point is $1$, this implies that the initial system must also be percolating (and that any percolation corrections correspond to errors from dropping higher order terms in the renormalized Hamiltonian). Similarly, a flow towards a non-percolating fixed point during which negative percolation corrections are attained implies that the starting point in coupling parameter space must be non-percolating. (Alternatively, the effects of percolation corrections on the confinement phase diagram can be analyzed by comparing to a site-bond percolation problem, see appendix \ref{sec:RG_appendix_sitebondperc}).

\begin{figure}
\centering
\leftline{a.}
{\includegraphics[width=0.45\textwidth]{\empty 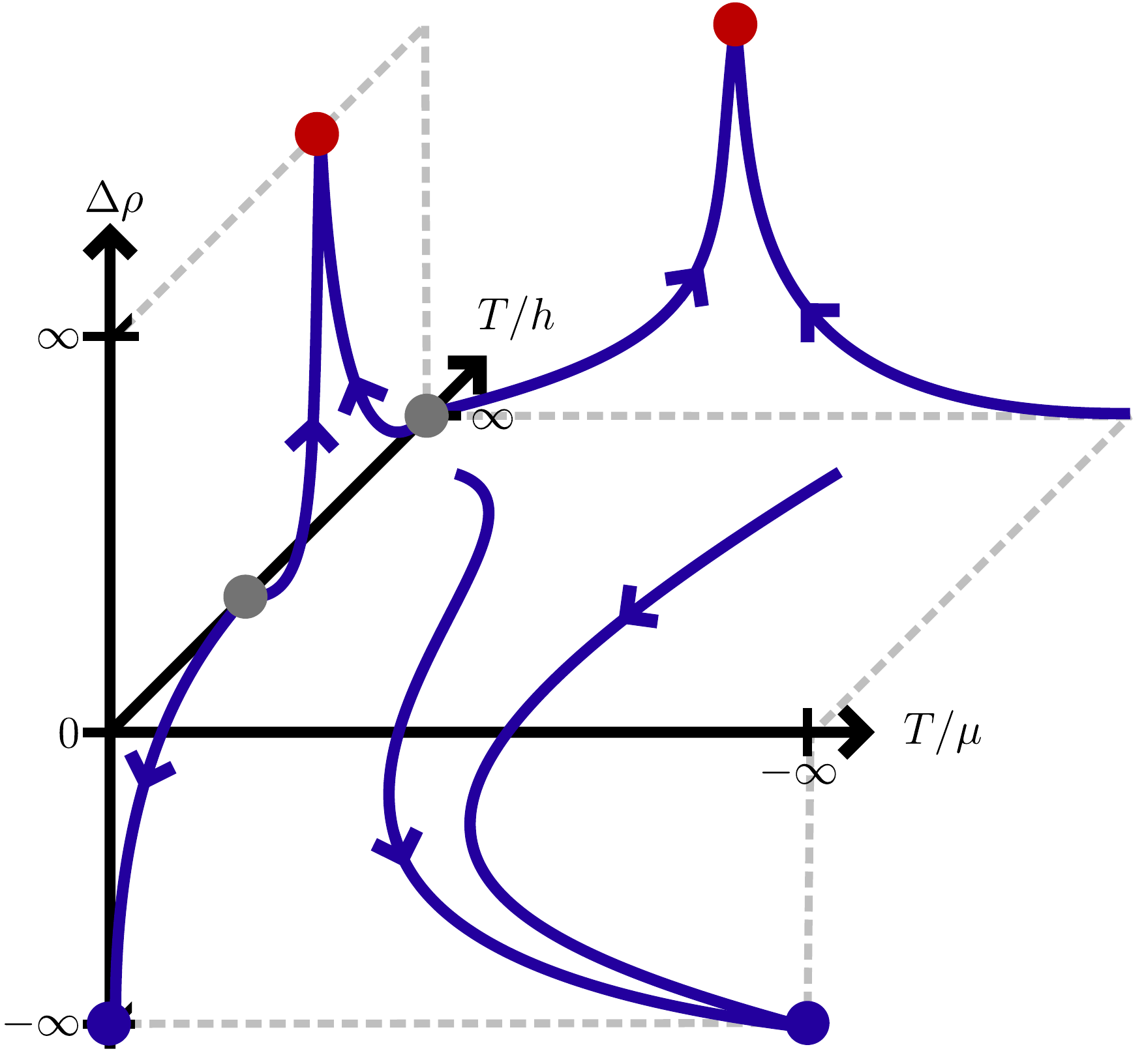}}
\leftline{b.}
{\includegraphics[width=0.45\textwidth]{\empty 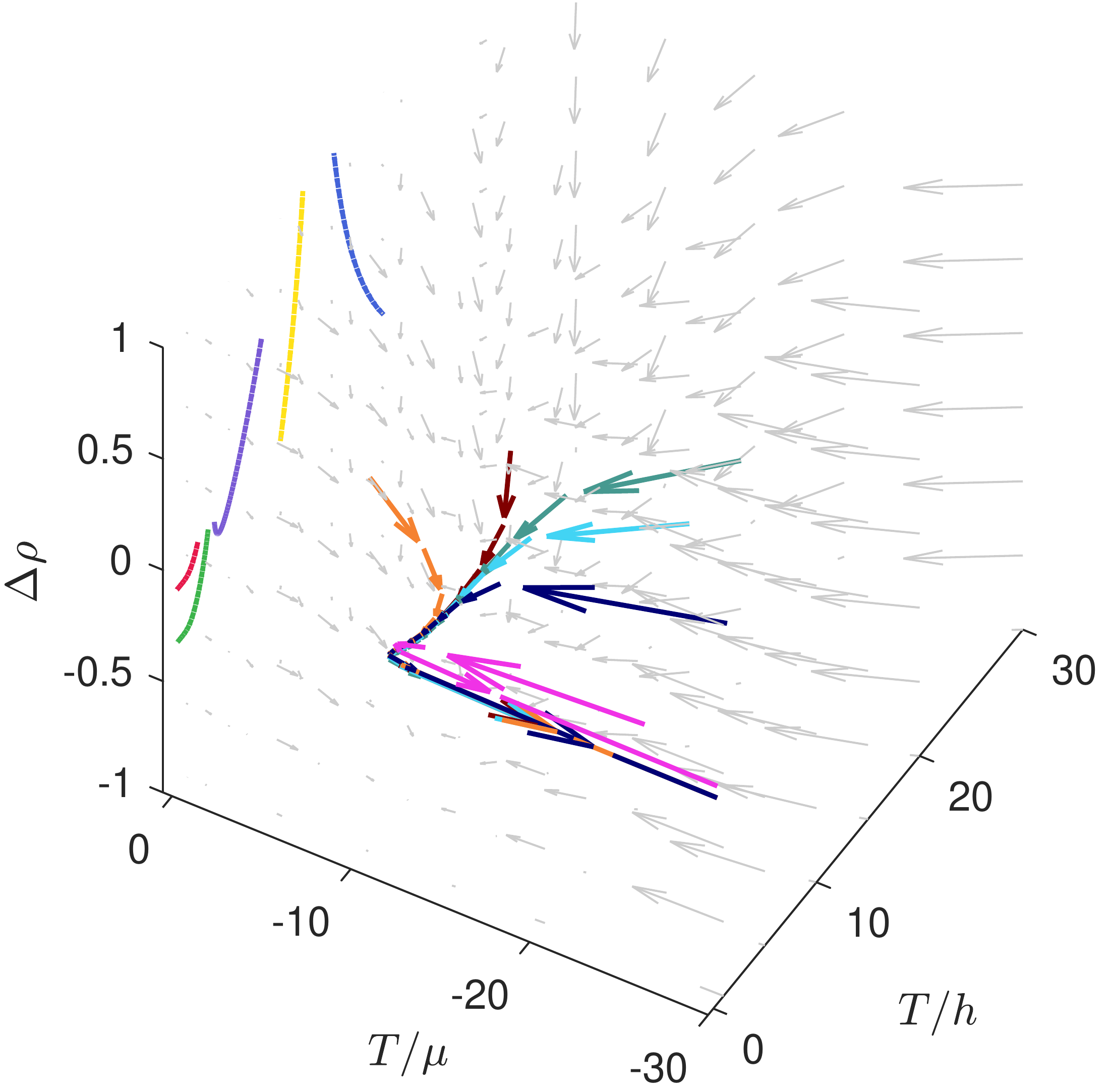}}
\caption{\label{fig:3Dflow} \textbf{Three-dimensional RG flow.} The flow of the coupling parameters and the percolation probability flow yield a three-dimensional combined RG flow, of which we show a schematic depiction (a) and a numerical plot (b). In the numerical plot, each arrow represents one RG step and the colours correspond to different starting points in parameter space. On the $T/\mu = 0$ axis the RG steps are too small to be visible. The percolation transition of the pure gauge theory with $T/\mu = 0$ is demonstrated by two different fixed points which the flow on this axis leads to.
}
\end{figure}

Recall that for all $T/h < \infty $ with nonzero $T / \mu$ the RG flow is towards a non-percolating fixed point. $\delta \rho$ is positive only in the region with $\mu / T \ll 0$ and $T / h$ large enough
(in Fig.~\ref{fig:PercCorrections} we see a point at $(T / \mu, T/h)  = (-1, 11)$ near the edge of this region).
The flow from any point $T/h < \infty $ with nonzero $T / \mu$ eventually leaves this region and $\Delta \rho$ is negative for any such parameter values. Thus we can conclude that the system does not percolate in these cases.

In the limit $T / h \to \infty$, the flow is towards a percolating fixed point. The percolation corrections then indicate that the system indeed percolates in this case for all $T / \mu$.

For $T / \mu = 0$ the percolation behavior changes with varying $T / h$. For large $T / h$, renormalizing the system significantly increases the percolation probability at each step. The percolation probability at these points must thus be greater than zero. Since Kolmogorov's zero-one law implies that the percolation probability in the thermodynamic limit is either zero or one \cite{Kolmogoroff1928, Grimmet}, the percolation probability in this region must be one. For small $T / h$, renormalizing leads to the non-percolating fixed point at $T / h = 0$ and the percolation corrections are negative. We thus conclude that there is a phase transition from a percolating system at large $T / h$ to a non-percolating system at small $T / h$.

The combined RG flow of the coupling parameters and the percolation probability can also be considered graphically as a three-dimensional flow (see Fig.~\ref{fig:3Dflow}). We see the percolation transition directly in the two stable fixed points of the flow for $T / \mu = 0$. These fixed points are in particular distinguished by the different levels occupied in terms of the percolation corrections accumulated during the flow there. Note that the fixed point $\delta \rho = 0$ of percolation probability flow is distinct from the fixed point of coupling parameter flow. 
However, a finite number of steps with negative percolation corrections at the beginning of the RG flow is outweighed by the infinite number of steps with positive percolation corrections that do not tend to zero in the RG flow toward the fixed point.
Thus the percolation transition of the pure gauge system occurs at the largest $T/h$ from which the flow ends in the region with positive $\delta \rho$.

\subsection{Phase diagram}

The results are summarized in the phase diagram shown in Fig.~\ref{fig:PhaseDiagram}a. The system exhibits confinement whenever $T/\mu < 0$ and $T/h < \infty$ and is deconfined in the limit $T/h \to \infty$. On the axis $T/|\mu| = 0$ (pure gauge theory), we find a percolation transition with confinement for small $T/h$ and deconfinement for large $T/h$. This is in good agreement with exact results and the predictions from Monte Carlo simulations \cite{Linsel2024}. For positive $\beta h$, we have in particular confirmed that \emph{any} finite chemical potential $\beta \mu$ (i.e. any nonzero particle density) results in a confined system with no percolating cluster of electric strings.

For the percolation transition of the pure gauge theory, the RG flow gives a critical $T / h$ value of $3.48$ which we compare to the exact value of $\sim2.27$ (the exact  value is identical to the Ising critical temperature). In the RG approach, higher order terms generated in the Hamiltonian are neglected at each step. It is thus expected that the RG results are qualitatively but not quantitatively correct.

\begin{figure}[t]
\centering
{\includegraphics[width=0.45\textwidth]{\empty 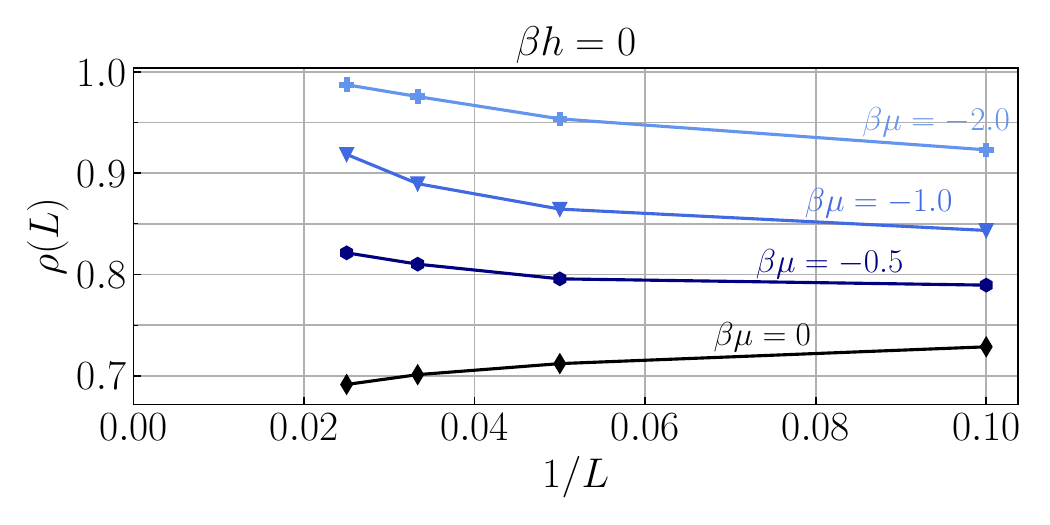}}
\caption{\label{fig:MC} \textbf{Monte Carlo at zero string tension.} We simulate Hamiltonian~(\ref{eqn:gcHam}) at $\beta h = 0$ (i.e. no string tension) using classical Monte Carlo. The error bars are too small to be visible. For $\beta \mu = 0$ it is exactly known that the system is percolating in the thermodynamic limit \cite{Linsel2024}. In this case we have $\rho(L \to \infty) \to 1/2$ since the system is \textit{directly at} the Bernoulli percolation threshold. For $\beta \mu < 0$ the percolation probability $\rho(L, \beta \mu)$ clearly increases with $L$ and $\rho(L, \beta \mu < 0) > \rho(L, \beta \mu = 0)$. We conjecture that $\rho(L \to \infty, \beta \mu < 0) \to 1$ due to Kolmogorov's zero-one law. These numerical results support our RG result that the system is percolating for $\beta h = 0$, $\beta \mu \leq 0$.}
\end{figure}



We perform Monte Carlo calculations to verify our results at zero string tension ($T / h \to \infty$), a regime not explored in previous studies \cite{Linsel2024}. In Fig.~\ref{fig:MC}, we show the percolation probability $\rho(\beta h = 0, L)$ for system sizes $L^2=10^2,20^2,30^2,40^2$ with open boundaries. The results agree with our renormalization group result that $(\beta h, \beta \mu) = (0,\beta \mu)$ always percolates.

Another way to analyze the RG flow by directly renormalizing the Monte-Carlo snapshots is presented in appendix \ref{sec:data_analysis}.

\section{Conclusion \& Outlook}

We have introduced a real-space RG scheme to probe confinement in $\mathbb{Z}_2$ LGTs using percolation probability as an order parameter. We successfully reproduced the confinement phase diagram of a classical $\mathbb{Z}_2$ LGT and found good agreement with exact and numerical benchmark results. In particular, we confirm the numerical result that a finite matter density prohibits a thermally deconfined phase on the 2D square lattice. In addition, we extended the phase diagram to regions not previously considered, including the case of vanishing string tension, and supported our RG results with Monte Carlo simulations.

The RG flow was considered as a simultaneous flow of the percolation probability alongside the coupling parameters. We envision that this RG ansatz can be generalized to other lattice geometries, to $\mathbb{Z}_N$ gauge symmetries, and to other models such as the Fradkin-Shenker model featuring quantum fluctuations in the future. We believe that our RG scheme has the potential to advance the analytical studies of confinement in $\mathbb{Z}_2$ LGTs and beyond.

\begin{acknowledgments}

We thank Lukas Homeier and Lode Pollet for insightful discussions.
This research was funded by the European Research Council (ERC) under the European Union’s Horizon 2020 research and innovation program (Grant Agreement no 948141) — ERC Starting Grant SimUcQuam, and by the Deutsche Forschungsgemeinschaft (DFG, German Research Foundation) under Germany's Excellence Strategy -- EXC-2111 -- 390814868 and via Research Unit FOR 2414 under project number 277974659.
This work was supported by the QuantERA grant DYNAMITE, by the Deutsche Forschungsgemeinschaft (DFG, German Research Foundation) under project number 499183856. This project was funded within the QuantERA II Programme that has received funding from the European Union’s Horizon 2020 research and innovation programme under Grand Agreement No 101017733.

\end{acknowledgments}


\bibliography{bibliography.bib} 

\newpage
\onecolumngrid
\appendix

\newpage

\section{Renormalization procedure - adjusted spanning cluster rule}

{\renewcommand{\arraystretch}{1}\begin{table*}[h!]
\begin{ruledtabular}
\begin{tabular}{c c c}
micro-configuration & Boltzmann weight & renormalized block configuration\\ 
\hline
    \subfloat{\centering\includegraphics[scale=0.2, valign=c]{\empty 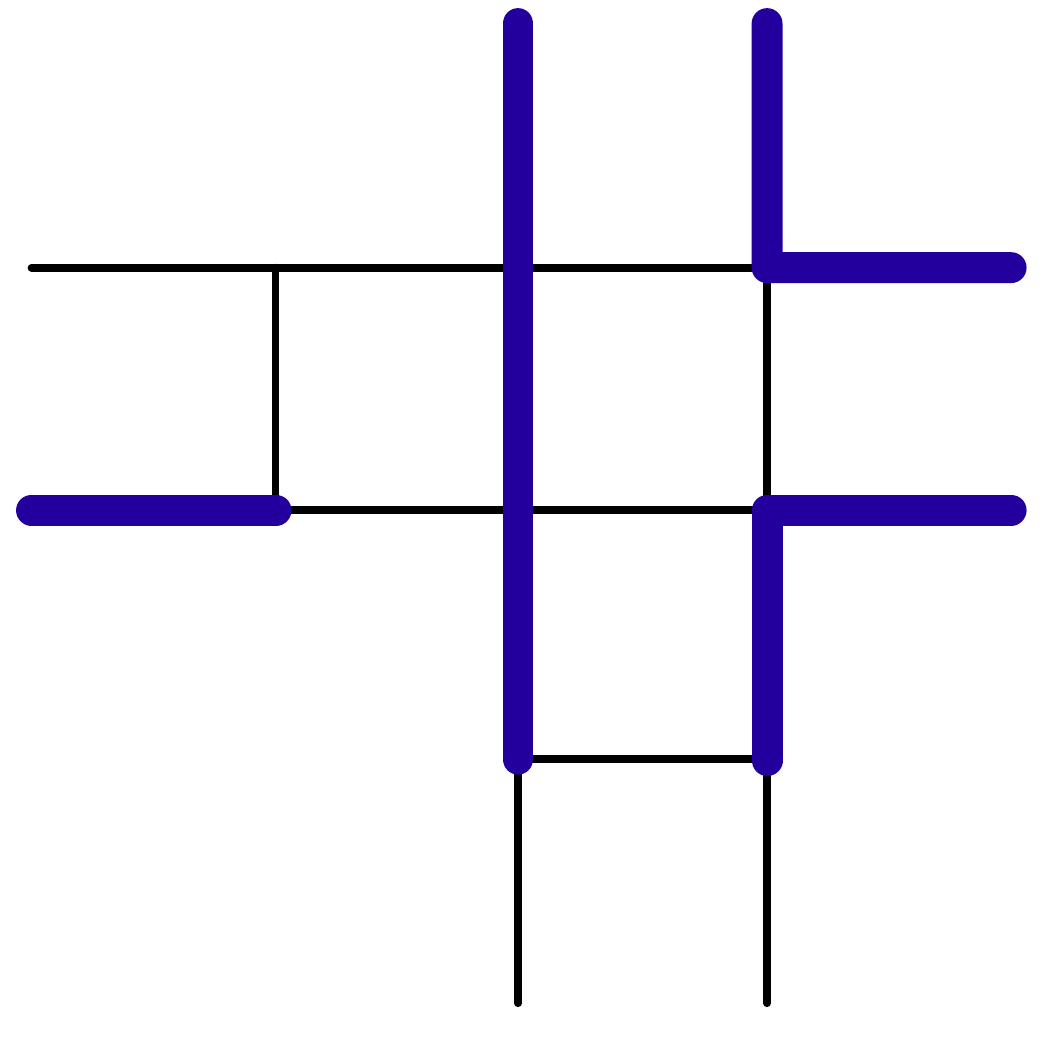}}
    & ${\mathrm{e}}^{2\,\beta \,h}$ &
    \subfloat{\centering\includegraphics[scale=0.2, valign=c]{\empty 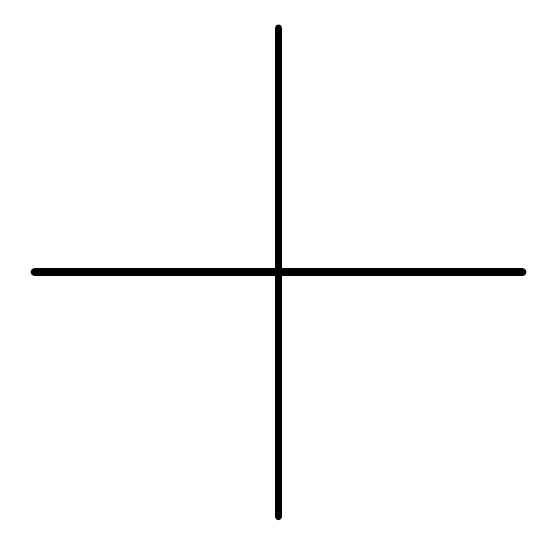}}
\\
    \subfloat{\centering\includegraphics[scale=0.2, valign=c]{\empty 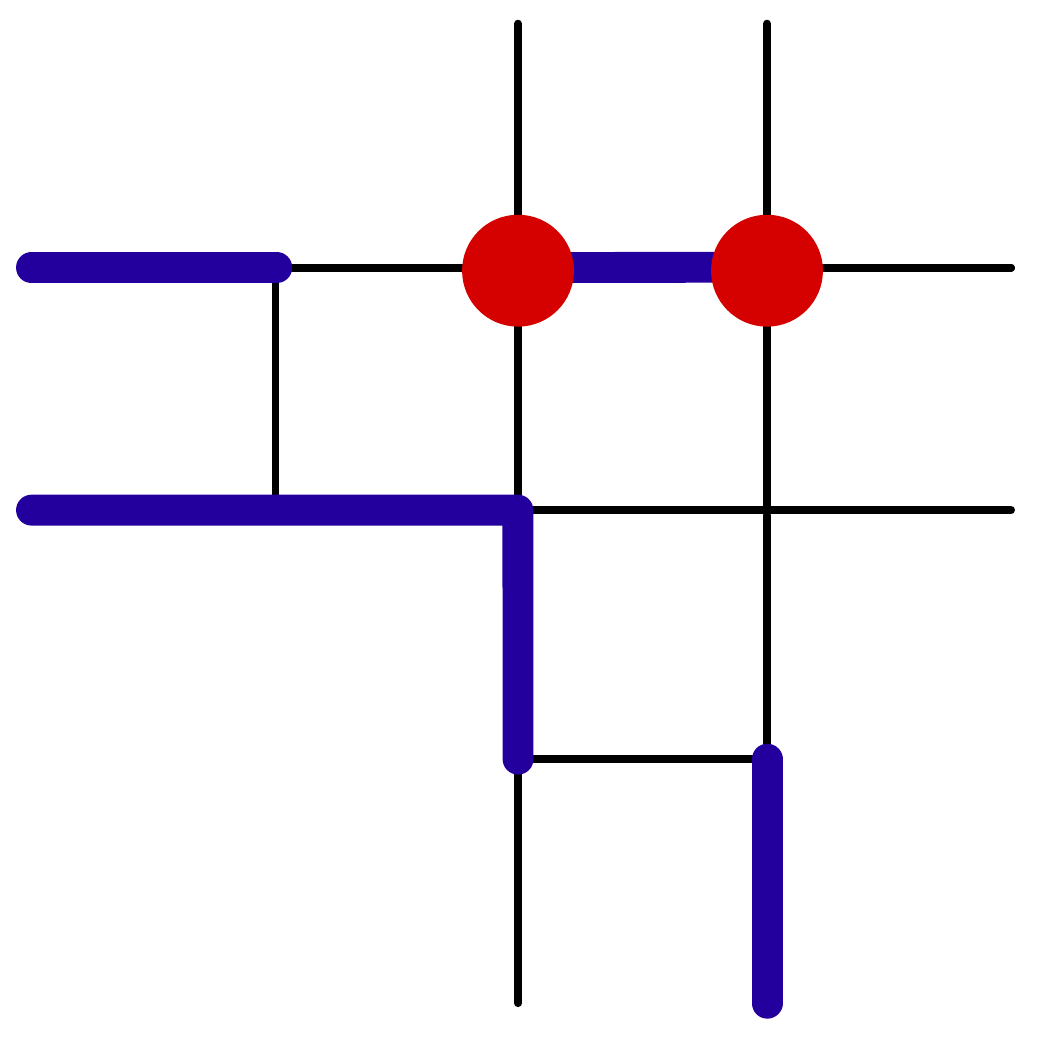}}
    & ${\mathrm{e}}^{6\,\beta \,h+2\,\beta \,\mu }$ &
    \subfloat{\centering\includegraphics[scale=0.2, valign=c]{\empty 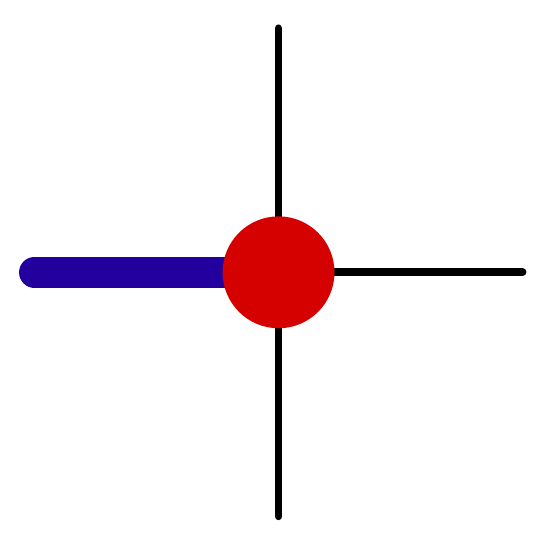}}
\\
    \subfloat{\centering\includegraphics[scale=0.2, valign=c]{\empty 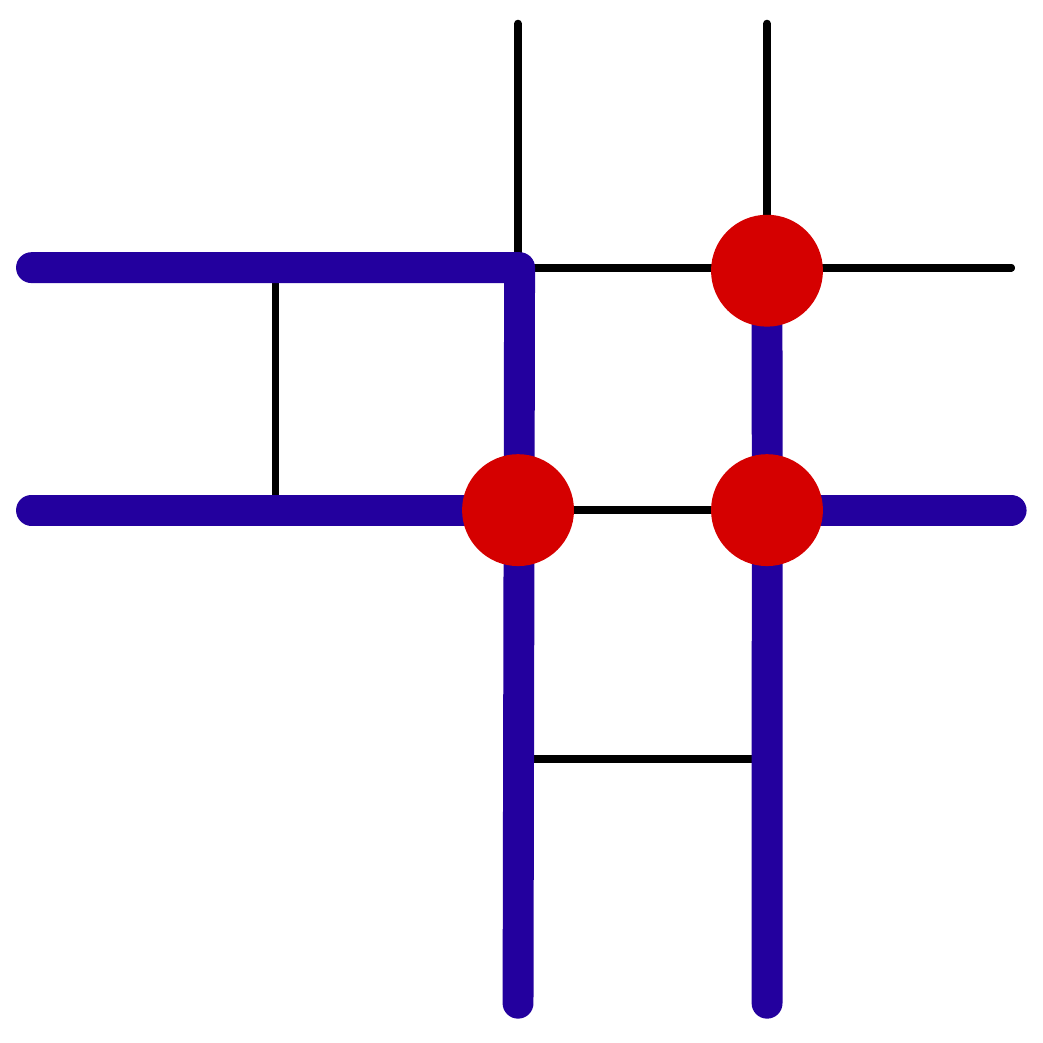}}
    & ${\mathrm{e}}^{3\,\beta \,\mu -4\,\beta \,h}$ &
    \subfloat{\centering\includegraphics[scale=0.2, valign=c]{\empty 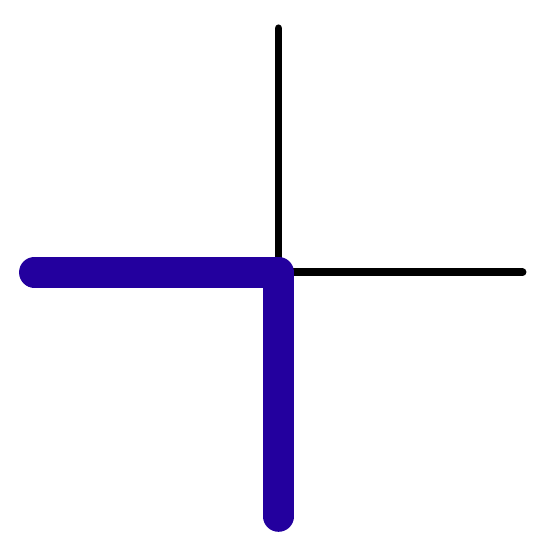}}
\\
    \subfloat{\centering\includegraphics[scale=0.2, valign=c]{\empty 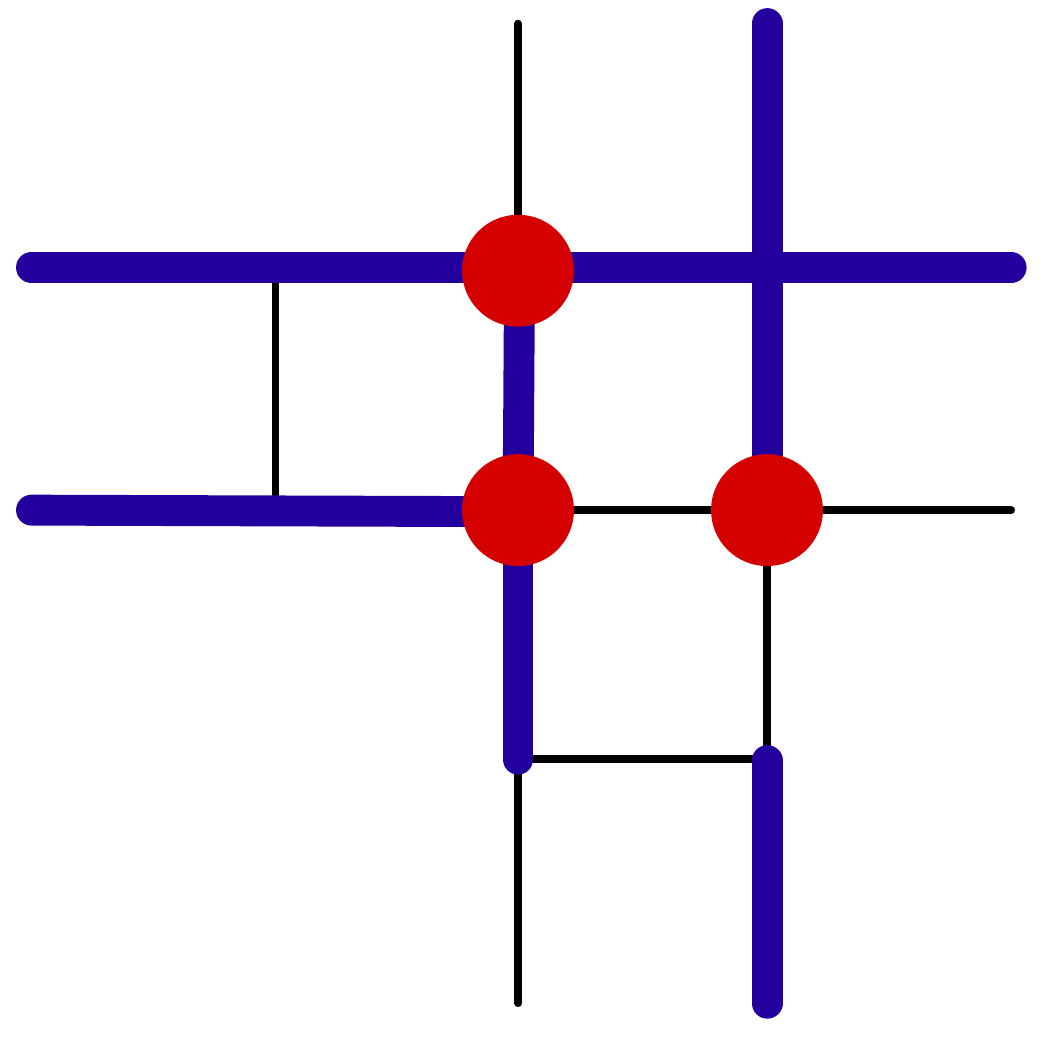}}
    & ${\mathrm{e}}^{3\,\beta \,\mu -4\,\beta \,h}$ &
    \subfloat{\centering\includegraphics[scale=0.2, valign=c]{\empty 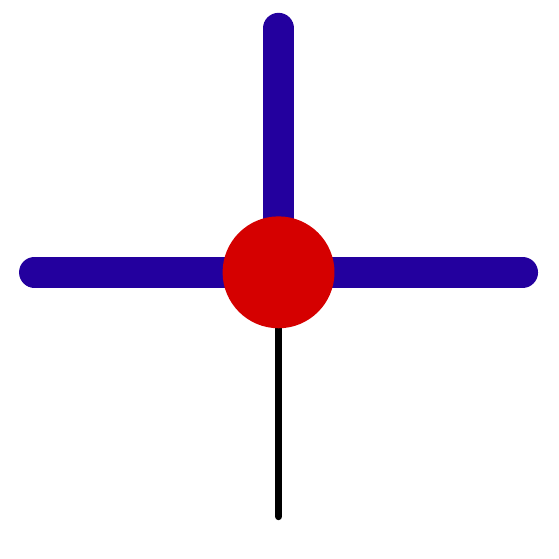}}
\\
    \subfloat{\centering\includegraphics[scale=0.2, valign=c]{\empty 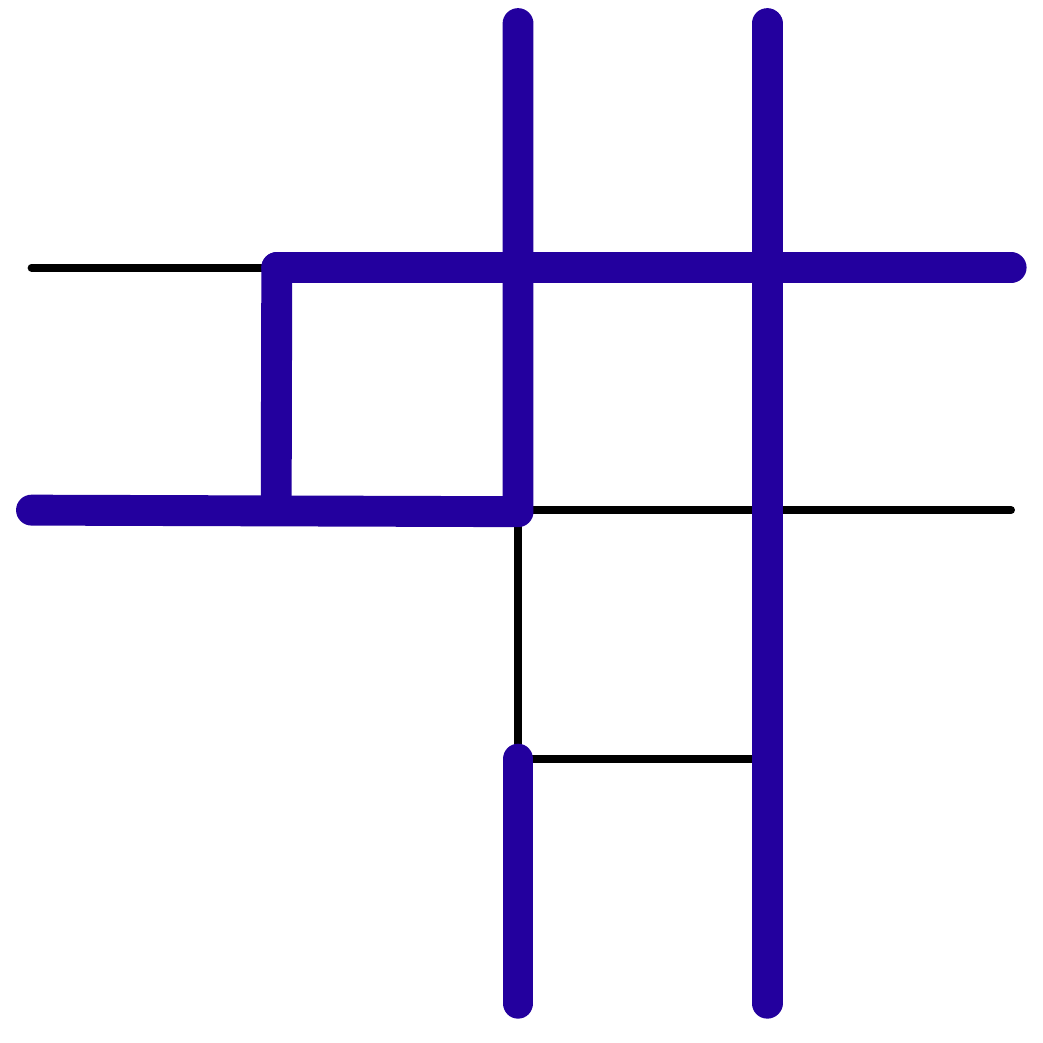}}
    & ${\mathrm{e}}^{-8\,\beta \,h}$ &
    \subfloat{\centering\includegraphics[scale=0.2, valign=c]{\empty 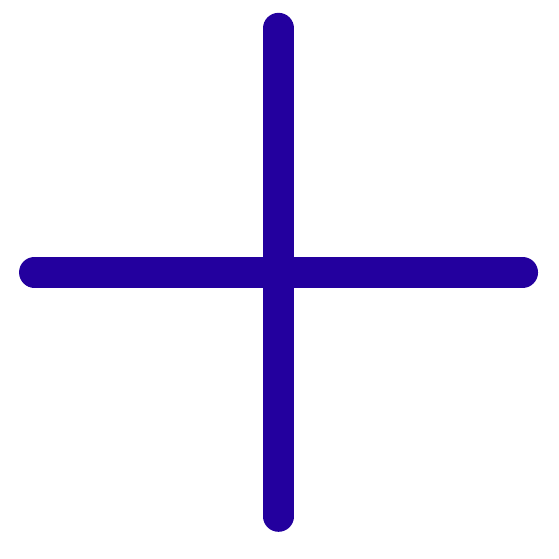}}
\\
\end{tabular}
\end{ruledtabular}
\caption{\label{tab:examples} \textbf{Adjusted spanning cluster rule.} We apply the adjusted spanning cluster renormalization to some example configurations.}
\end{table*}}

\section{Details of RG flow} \label{sec:RG_appendix_flowderiv}

To determine the renormalized Hamiltonian $\hat{H}'$, we classify all allowed micro-configurations $\{\hat{\tau}^x_1, ..., \hat{\tau}^x_{18}\}$ of a block according to the assigned block configuration $\{\hat{\tau}^x_\text{I}, \hat{\tau}^x_\text{II}, \hat{\tau}^x_\text{III} \hat{\tau}^x_\text{IV}\}$. Taking a specific block-state and adding the Boltzmann weights of all micro-states which are renormalized to this gives the expectation value
\begin{equation}
\exp(- \beta \hat{H}'\left(\left\{\hat{\tau}^x_\text{I}, \hat{\tau}^x_\text{II}, \hat{\tau}^x_\text{III} \hat{\tau}^x_\text{IV}\right\}\right) )
=  \sum_{\text{resp. microstates}} 
\exp(-\beta \hat{H}\left(\left\{\hat{\tau}^x_{1}, ..., \hat{\tau}^x_{18}\right\}, h, \mu\right)).
\end{equation} 

Using symbolic computation to calculate these sums exactly, we obtain the following results:

\begin{eqnarray*}
c \cdot &&e^{4 \beta h'} \overset{!}{=} e^{- \beta \hat{H}'(1,1,1,1)}
\\ = && 293\,{\mathrm{e}}^{-2\,\beta \,h} +1165\,{\mathrm{e}}^{2\,\beta \,h} +62\,{\mathrm{e}}^{-4\,\beta \,h} +1160\,{\mathrm{e}}^{4\,\beta \,h} +6\,{\mathrm{e}}^{-6\,\beta \,h} +815\,{\mathrm{e}}^{6\,\beta \,h} +442\,{\mathrm{e}}^{8\,\beta \,h} +185\,{\mathrm{e}}^{10\,\beta \,h} +60\,{\mathrm{e}}^{12\,\beta \,h} 
\\ && +19\,{\mathrm{e}}^{14\,\beta \,h} +6\,{\mathrm{e}}^{16\,\beta \,h} +{\mathrm{e}}^{18\,\beta \,h} +124\,{\mathrm{e}}^{\beta \,\mu } +158\,{\mathrm{e}}^{2\,\beta \,\mu } +68\,{\mathrm{e}}^{3\,\beta \,\mu } +8\,{\mathrm{e}}^{4\,\beta \,\mu } +12\,{\mathrm{e}}^{-2\,\beta \,h} \,{\mathrm{e}}^{\beta \,\mu } +480\,{\mathrm{e}}^{2\,\beta \,h} \,{\mathrm{e}}^{\beta \,\mu } 
\\ && +15\,{\mathrm{e}}^{-2\,\beta \,h} \,{\mathrm{e}}^{2\,\beta \,\mu } +663\,{\mathrm{e}}^{2\,\beta \,h} \,{\mathrm{e}}^{2\,\beta \,\mu } +900\,{\mathrm{e}}^{4\,\beta \,h} \,{\mathrm{e}}^{\beta \,\mu } +4\,{\mathrm{e}}^{-2\,\beta \,h} \,{\mathrm{e}}^{3\,\beta \,\mu } +392\,{\mathrm{e}}^{2\,\beta \,h} \,{\mathrm{e}}^{3\,\beta \,\mu } +80\,{\mathrm{e}}^{2\,\beta \,h} \,{\mathrm{e}}^{4\,\beta \,\mu } +1464\,{\mathrm{e}}^{4\,\beta \,h} \,{\mathrm{e}}^{2\,\beta \,\mu } 
\\ && +1032\,{\mathrm{e}}^{6\,\beta \,h} \,{\mathrm{e}}^{\beta \,\mu } +1044\,{\mathrm{e}}^{4\,\beta \,h} \,{\mathrm{e}}^{3\,\beta \,\mu } +278\,{\mathrm{e}}^{4\,\beta \,h} \,{\mathrm{e}}^{4\,\beta \,\mu } +1858\,{\mathrm{e}}^{6\,\beta \,h} \,{\mathrm{e}}^{2\,\beta \,\mu } +808\,{\mathrm{e}}^{8\,\beta \,h} \,{\mathrm{e}}^{\beta \,\mu } +1456\,{\mathrm{e}}^{6\,\beta \,h} \,{\mathrm{e}}^{3\,\beta \,\mu } 
\\ &&  +424\,{\mathrm{e}}^{6\,\beta \,h} \,{\mathrm{e}}^{4\,\beta \,\mu } +1484\,{\mathrm{e}}^{8\,\beta \,h} \,{\mathrm{e}}^{2\,\beta \,\mu } +456\,{\mathrm{e}}^{10\,\beta \,h} \,{\mathrm{e}}^{\beta \,\mu } +1168\,{\mathrm{e}}^{8\,\beta \,h} \,{\mathrm{e}}^{3\,\beta \,\mu } +332\,{\mathrm{e}}^{8\,\beta \,h} \,{\mathrm{e}}^{4\,\beta \,\mu } +784\,{\mathrm{e}}^{10\,\beta \,h} \,{\mathrm{e}}^{2\,\beta \,\mu } 
\\ && +192\,{\mathrm{e}}^{12\,\beta \,h} \,{\mathrm{e}}^{\beta \,\mu } +552\,{\mathrm{e}}^{10\,\beta \,h} \,{\mathrm{e}}^{3\,\beta \,\mu } +138\,{\mathrm{e}}^{10\,\beta \,h} \,{\mathrm{e}}^{4\,\beta \,\mu } +268\,{\mathrm{e}}^{12\,\beta \,h} \,{\mathrm{e}}^{2\,\beta \,\mu } +56\,{\mathrm{e}}^{14\,\beta \,h} \,{\mathrm{e}}^{\beta \,\mu } +144\,{\mathrm{e}}^{12\,\beta \,h} \,{\mathrm{e}}^{3\,\beta \,\mu } 
\\ && +28\,{\mathrm{e}}^{12\,\beta \,h} \,{\mathrm{e}}^{4\,\beta \,\mu } +52\,{\mathrm{e}}^{14\,\beta \,h} \,{\mathrm{e}}^{2\,\beta \,\mu } +8\,{\mathrm{e}}^{16\,\beta \,h} \,{\mathrm{e}}^{\beta \,\mu } +16\,{\mathrm{e}}^{14\,\beta \,h} \,{\mathrm{e}}^{3\,\beta \,\mu } +2\,{\mathrm{e}}^{14\,\beta \,h} \,{\mathrm{e}}^{4\,\beta \,\mu } +4\,{\mathrm{e}}^{16\,\beta \,h} \,{\mathrm{e}}^{2\,\beta \,\mu } +754
\end{eqnarray*}

\begin{eqnarray*}
c \cdot &&e^{2 \beta h'} e^{\beta \mu'} \overset{!}{=} e^{- \beta \hat{H'}(-1,1,1,1)} = e^{- \beta \hat{H'}(1,-1,1,1)} = 
e^{- \beta \hat{H'}(1,1,-1,1)} = e^{- \beta \hat{H'}(1,1,1,-1)}
\\ = &&
578\,{\mathrm{e}}^{\beta \,\mu } +840\,{\mathrm{e}}^{2\,\beta \,\mu } +542\,{\mathrm{e}}^{3\,\beta \,\mu } +133\,{\mathrm{e}}^{4\,\beta \,\mu } +174\,{\mathrm{e}}^{-2\,\beta \,h} \,{\mathrm{e}}^{\beta \,\mu } +1002\,{\mathrm{e}}^{2\,\beta \,h} \,{\mathrm{e}}^{\beta \,\mu } +277\,{\mathrm{e}}^{-2\,\beta \,h} \,{\mathrm{e}}^{2\,\beta \,\mu } +1454\,{\mathrm{e}}^{2\,\beta \,h} \,{\mathrm{e}}^{2\,\beta \,\mu } 
\\ && +26\,{\mathrm{e}}^{-4\,\beta \,h} \,{\mathrm{e}}^{\beta \,\mu } +1044\,{\mathrm{e}}^{4\,\beta \,h} \,{\mathrm{e}}^{\beta \,\mu } +186\,{\mathrm{e}}^{-2\,\beta \,h} \,{\mathrm{e}}^{3\,\beta \,\mu } +958\,{\mathrm{e}}^{2\,\beta \,h} \,{\mathrm{e}}^{3\,\beta \,\mu } +45\,{\mathrm{e}}^{-2\,\beta \,h} \,{\mathrm{e}}^{4\,\beta \,\mu } +234\,{\mathrm{e}}^{2\,\beta \,h} \,{\mathrm{e}}^{4\,\beta \,\mu } 
\\ && +44\,{\mathrm{e}}^{-4\,\beta \,h} \,{\mathrm{e}}^{2\,\beta \,\mu } +1502\,{\mathrm{e}}^{4\,\beta \,h} \,{\mathrm{e}}^{2\,\beta \,\mu } +702\,{\mathrm{e}}^{6\,\beta \,h} \,{\mathrm{e}}^{\beta \,\mu } +38\,{\mathrm{e}}^{-4\,\beta \,h} \,{\mathrm{e}}^{3\,\beta \,\mu } +972\,{\mathrm{e}}^{4\,\beta \,h} \,{\mathrm{e}}^{3\,\beta \,\mu } +9\,{\mathrm{e}}^{-4\,\beta \,h} \,{\mathrm{e}}^{4\,\beta \,\mu } 
\\ && +239\,{\mathrm{e}}^{4\,\beta \,h} \,{\mathrm{e}}^{4\,\beta \,\mu } +6\,{\mathrm{e}}^{-6\,\beta \,h} \,{\mathrm{e}}^{2\,\beta \,\mu } +948\,{\mathrm{e}}^{6\,\beta \,h} \,{\mathrm{e}}^{2\,\beta \,\mu } +304\,{\mathrm{e}}^{8\,\beta \,h} \,{\mathrm{e}}^{\beta \,\mu } +584\,{\mathrm{e}}^{6\,\beta \,h} \,{\mathrm{e}}^{3\,\beta \,\mu } +2\,{\mathrm{e}}^{-6\,\beta \,h} \,{\mathrm{e}}^{4\,\beta \,\mu } +141\,{\mathrm{e}}^{6\,\beta \,h} \,{\mathrm{e}}^{4\,\beta \,\mu } 
\\ && +370\,{\mathrm{e}}^{8\,\beta \,h} \,{\mathrm{e}}^{2\,\beta \,\mu } +84\,{\mathrm{e}}^{10\,\beta \,h} \,{\mathrm{e}}^{\beta \,\mu } +212\,{\mathrm{e}}^{8\,\beta \,h} \,{\mathrm{e}}^{3\,\beta \,\mu } +45\,{\mathrm{e}}^{8\,\beta \,h} \,{\mathrm{e}}^{4\,\beta \,\mu } +87\,{\mathrm{e}}^{10\,\beta \,h} \,{\mathrm{e}}^{2\,\beta \,\mu } +16\,{\mathrm{e}}^{12\,\beta \,h} \,{\mathrm{e}}^{\beta \,\mu } +44\,{\mathrm{e}}^{10\,\beta \,h} \,{\mathrm{e}}^{3\,\beta \,\mu } 
\\ && +6\,{\mathrm{e}}^{10\,\beta \,h} \,{\mathrm{e}}^{4\,\beta \,\mu } +10\,{\mathrm{e}}^{12\,\beta \,h} \,{\mathrm{e}}^{2\,\beta \,\mu } +2\,{\mathrm{e}}^{14\,\beta \,h} \,{\mathrm{e}}^{\beta \,\mu } +4\,{\mathrm{e}}^{12\,\beta \,h} \,{\mathrm{e}}^{3\,\beta \,\mu }
\end{eqnarray*}

\begin{eqnarray*}
c  &&\overset{!}{=} e^{- \beta \hat{H'}(-1,-1,1,1)} = e^{- \beta \hat{H'}(1,-1,-1,1)} = 
e^{- \beta \hat{H'}(1,1,-1,-1)} = e^{- \beta \hat{H'}(-1,1,1,-1)}
\\ = &&
232\,{\mathrm{e}}^{-2\,\beta \,h} +266\,{\mathrm{e}}^{2\,\beta \,h} +104\,{\mathrm{e}}^{-4\,\beta \,h} +146\,{\mathrm{e}}^{4\,\beta \,h} +25\,{\mathrm{e}}^{-6\,\beta \,h} +54\,{\mathrm{e}}^{6\,\beta \,h} +4\,{\mathrm{e}}^{-8\,\beta \,h} +14\,{\mathrm{e}}^{8\,\beta \,h} +{\mathrm{e}}^{-10\,\beta \,h} +2\,{\mathrm{e}}^{10\,\beta \,h} 
\\ && +1224\,{\mathrm{e}}^{\beta \,\mu } +1846\,{\mathrm{e}}^{2\,\beta \,\mu } +1280\,{\mathrm{e}}^{3\,\beta \,\mu } +332\,{\mathrm{e}}^{4\,\beta \,\mu } +894\,{\mathrm{e}}^{-2\,\beta \,h} \,{\mathrm{e}}^{\beta \,\mu } +1030\,{\mathrm{e}}^{2\,\beta \,h} \,{\mathrm{e}}^{\beta \,\mu } +1343\,{\mathrm{e}}^{-2\,\beta \,h} \,{\mathrm{e}}^{2\,\beta \,\mu } 
\\ && +1597\,{\mathrm{e}}^{2\,\beta \,h} \,{\mathrm{e}}^{2\,\beta \,\mu } +392\,{\mathrm{e}}^{-4\,\beta \,h} \,{\mathrm{e}}^{\beta \,\mu } +560\,{\mathrm{e}}^{4\,\beta \,h} \,{\mathrm{e}}^{\beta \,\mu } +934\,{\mathrm{e}}^{-2\,\beta \,h} \,{\mathrm{e}}^{3\,\beta \,\mu } +1034\,{\mathrm{e}}^{2\,\beta \,h} \,{\mathrm{e}}^{3\,\beta \,\mu } +258\,{\mathrm{e}}^{-2\,\beta \,h} \,{\mathrm{e}}^{4\,\beta \,\mu } 
\\ && +246\,{\mathrm{e}}^{2\,\beta \,h} \,{\mathrm{e}}^{4\,\beta \,\mu } +616\,{\mathrm{e}}^{-4\,\beta \,h} \,{\mathrm{e}}^{2\,\beta \,\mu } +836\,{\mathrm{e}}^{4\,\beta \,h} \,{\mathrm{e}}^{2\,\beta \,\mu } +118\,{\mathrm{e}}^{-6\,\beta \,h} \,{\mathrm{e}}^{\beta \,\mu } +186\,{\mathrm{e}}^{6\,\beta \,h} \,{\mathrm{e}}^{\beta \,\mu } +416\,{\mathrm{e}}^{-4\,\beta \,h} \,{\mathrm{e}}^{3\,\beta \,\mu } 
\\ && +504\,{\mathrm{e}}^{4\,\beta \,h} \,{\mathrm{e}}^{3\,\beta \,\mu } +108\,{\mathrm{e}}^{-4\,\beta \,h} \,{\mathrm{e}}^{4\,\beta \,\mu } +106\,{\mathrm{e}}^{4\,\beta \,h} \,{\mathrm{e}}^{4\,\beta \,\mu } +159\,{\mathrm{e}}^{-6\,\beta \,h} \,{\mathrm{e}}^{2\,\beta \,\mu } +263\,{\mathrm{e}}^{6\,\beta \,h} \,{\mathrm{e}}^{2\,\beta \,\mu } +20\,{\mathrm{e}}^{-8\,\beta \,h} \,{\mathrm{e}}^{\beta \,\mu } 
\\ && +28\,{\mathrm{e}}^{8\,\beta \,h} \,{\mathrm{e}}^{\beta \,\mu } +114\,{\mathrm{e}}^{-6\,\beta \,h} \,{\mathrm{e}}^{3\,\beta \,\mu } +146\,{\mathrm{e}}^{6\,\beta \,h} \,{\mathrm{e}}^{3\,\beta \,\mu } +20\,{\mathrm{e}}^{-6\,\beta \,h} \,{\mathrm{e}}^{4\,\beta \,\mu } +24\,{\mathrm{e}}^{6\,\beta \,h} \,{\mathrm{e}}^{4\,\beta \,\mu } +26\,{\mathrm{e}}^{-8\,\beta \,h} \,{\mathrm{e}}^{2\,\beta \,\mu } 
\\ && +54\,{\mathrm{e}}^{8\,\beta \,h} \,{\mathrm{e}}^{2\,\beta \,\mu } +12\,{\mathrm{e}}^{-8\,\beta \,h} \,{\mathrm{e}}^{3\,\beta \,\mu } +20\,{\mathrm{e}}^{8\,\beta \,h} \,{\mathrm{e}}^{3\,\beta \,\mu } +2\,{\mathrm{e}}^{-8\,\beta \,h} \,{\mathrm{e}}^{4\,\beta \,\mu } +2\,{\mathrm{e}}^{8\,\beta \,h} \,{\mathrm{e}}^{4\,\beta \,\mu } +3\,{\mathrm{e}}^{-10\,\beta \,h} \,{\mathrm{e}}^{2\,\beta \,\mu } 
\\ && +7\,{\mathrm{e}}^{10\,\beta \,h} \,{\mathrm{e}}^{2\,\beta \,\mu } +312
\end{eqnarray*}

\begin{eqnarray*}
c  &&\overset{!}{=} e^{- \beta \hat{H'}(-1,1,-1,1)} = e^{- \beta \hat{H'}(1,-1,1,-1)}
\\ = &&
219\,{\mathrm{e}}^{-2\,\beta \,h} +261\,{\mathrm{e}}^{2\,\beta \,h} +95\,{\mathrm{e}}^{-4\,\beta \,h} +149\,{\mathrm{e}}^{4\,\beta \,h} +22\,{\mathrm{e}}^{-6\,\beta \,h} +60\,{\mathrm{e}}^{6\,\beta \,h} +2\,{\mathrm{e}}^{-8\,\beta \,h} +16\,{\mathrm{e}}^{8\,\beta \,h} +2\,{\mathrm{e}}^{10\,\beta \,h} +1116\,{\mathrm{e}}^{\beta \,\mu } 
\\ && +1604\,{\mathrm{e}}^{2\,\beta \,\mu } +992\,{\mathrm{e}}^{3\,\beta \,\mu } +210\,{\mathrm{e}}^{4\,\beta \,\mu } +849\,{\mathrm{e}}^{-2\,\beta \,h} \,{\mathrm{e}}^{\beta \,\mu } +903\,{\mathrm{e}}^{2\,\beta \,h} \,{\mathrm{e}}^{\beta \,\mu } +1241\,{\mathrm{e}}^{-2\,\beta \,h} \,{\mathrm{e}}^{2\,\beta \,\mu } +1242\,{\mathrm{e}}^{2\,\beta \,h} \,{\mathrm{e}}^{2\,\beta \,\mu } 
\\ && +368\,{\mathrm{e}}^{-4\,\beta \,h} \,{\mathrm{e}}^{\beta \,\mu } +464\,{\mathrm{e}}^{4\,\beta \,h} \,{\mathrm{e}}^{\beta \,\mu } +789\,{\mathrm{e}}^{-2\,\beta \,h} \,{\mathrm{e}}^{3\,\beta \,\mu } +721\,{\mathrm{e}}^{2\,\beta \,h} \,{\mathrm{e}}^{3\,\beta \,\mu } +182\,{\mathrm{e}}^{-2\,\beta \,h} \,{\mathrm{e}}^{4\,\beta \,\mu } +153\,{\mathrm{e}}^{2\,\beta \,h} \,{\mathrm{e}}^{4\,\beta \,\mu } 
\\ && +570\,{\mathrm{e}}^{-4\,\beta \,h} \,{\mathrm{e}}^{2\,\beta \,\mu } +564\,{\mathrm{e}}^{4\,\beta \,h} \,{\mathrm{e}}^{2\,\beta \,\mu } +89\,{\mathrm{e}}^{-6\,\beta \,h} \,{\mathrm{e}}^{\beta \,\mu } +139\,{\mathrm{e}}^{6\,\beta \,h} \,{\mathrm{e}}^{\beta \,\mu } +400\,{\mathrm{e}}^{-4\,\beta \,h} \,{\mathrm{e}}^{3\,\beta \,\mu } +292\,{\mathrm{e}}^{4\,\beta \,h} \,{\mathrm{e}}^{3\,\beta \,\mu } 
\\ && +105\,{\mathrm{e}}^{-4\,\beta \,h} \,{\mathrm{e}}^{4\,\beta \,\mu } +69\,{\mathrm{e}}^{4\,\beta \,h} \,{\mathrm{e}}^{4\,\beta \,\mu } +139\,{\mathrm{e}}^{-6\,\beta \,h} \,{\mathrm{e}}^{2\,\beta \,\mu } +146\,{\mathrm{e}}^{6\,\beta \,h} \,{\mathrm{e}}^{2\,\beta \,\mu } +10\,{\mathrm{e}}^{-8\,\beta \,h} \,{\mathrm{e}}^{\beta \,\mu } +18\,{\mathrm{e}}^{8\,\beta \,h} \,{\mathrm{e}}^{\beta \,\mu } 
\\ && +135\,{\mathrm{e}}^{-6\,\beta \,h} \,{\mathrm{e}}^{3\,\beta \,\mu } +63\,{\mathrm{e}}^{6\,\beta \,h} \,{\mathrm{e}}^{3\,\beta \,\mu } +41\,{\mathrm{e}}^{-6\,\beta \,h} \,{\mathrm{e}}^{4\,\beta \,\mu } +18\,{\mathrm{e}}^{6\,\beta \,h} \,{\mathrm{e}}^{4\,\beta \,\mu } +18\,{\mathrm{e}}^{-8\,\beta \,h} \,{\mathrm{e}}^{2\,\beta \,\mu } +22\,{\mathrm{e}}^{8\,\beta \,h} \,{\mathrm{e}}^{2\,\beta \,\mu } 
\\ && +22\,{\mathrm{e}}^{-8\,\beta \,h} \,{\mathrm{e}}^{3\,\beta \,\mu } +6\,{\mathrm{e}}^{8\,\beta \,h} \,{\mathrm{e}}^{3\,\beta \,\mu } +12\,{\mathrm{e}}^{-8\,\beta \,h} \,{\mathrm{e}}^{4\,\beta \,\mu } +2\,{\mathrm{e}}^{8\,\beta \,h} \,{\mathrm{e}}^{4\,\beta \,\mu } +2\,{\mathrm{e}}^{-10\,\beta \,h} \,{\mathrm{e}}^{2\,\beta \,\mu } +2\,{\mathrm{e}}^{10\,\beta \,h} \,{\mathrm{e}}^{2\,\beta \,\mu } 
\\ && +2\,{\mathrm{e}}^{-10\,\beta \,h} \,{\mathrm{e}}^{4\,\beta \,\mu } +302
\end{eqnarray*}

\begin{eqnarray*}
c \cdot &&e^{-2 \beta h'} e^{\beta \mu'}  \overset{!}{=} e^{- \beta \hat{H'}(-1,-1,-1,1)} = e^{- \beta \hat{H'}(-1,-1,1,-1)} = 
e^{- \beta \hat{H'}(-1,1,-1,-1)} = e^{- \beta \hat{H'}(1,-1,-1,-1)}
\\ = &&
732\,{\mathrm{e}}^{\beta \,\mu } +1166\,{\mathrm{e}}^{2\,\beta \,\mu } +760\,{\mathrm{e}}^{3\,\beta \,\mu } +182\,{\mathrm{e}}^{4\,\beta \,\mu } +1161\,{\mathrm{e}}^{-2\,\beta \,h} \,{\mathrm{e}}^{\beta \,\mu } +301\,{\mathrm{e}}^{2\,\beta \,h} \,{\mathrm{e}}^{\beta \,\mu } +1785\,{\mathrm{e}}^{-2\,\beta \,h} \,{\mathrm{e}}^{2\,\beta \,\mu } +420\,{\mathrm{e}}^{2\,\beta \,h} \,{\mathrm{e}}^{2\,\beta \,\mu } 
\\ && +1116\,{\mathrm{e}}^{-4\,\beta \,h} \,{\mathrm{e}}^{\beta \,\mu } +82\,{\mathrm{e}}^{4\,\beta \,h} \,{\mathrm{e}}^{\beta \,\mu } +1259\,{\mathrm{e}}^{-2\,\beta \,h} \,{\mathrm{e}}^{3\,\beta \,\mu } +267\,{\mathrm{e}}^{2\,\beta \,h} \,{\mathrm{e}}^{3\,\beta \,\mu } +325\,{\mathrm{e}}^{-2\,\beta \,h} \,{\mathrm{e}}^{4\,\beta \,\mu } +57\,{\mathrm{e}}^{2\,\beta \,h} \,{\mathrm{e}}^{4\,\beta \,\mu } 
\\ && +1720\,{\mathrm{e}}^{-4\,\beta \,h} \,{\mathrm{e}}^{2\,\beta \,\mu } +62\,{\mathrm{e}}^{4\,\beta \,h} \,{\mathrm{e}}^{2\,\beta \,\mu } +657\,{\mathrm{e}}^{-6\,\beta \,h} \,{\mathrm{e}}^{\beta \,\mu } +11\,{\mathrm{e}}^{6\,\beta \,h} \,{\mathrm{e}}^{\beta \,\mu } +1200\,{\mathrm{e}}^{-4\,\beta \,h} \,{\mathrm{e}}^{3\,\beta \,\mu } +58\,{\mathrm{e}}^{4\,\beta \,h} \,{\mathrm{e}}^{3\,\beta \,\mu } 
\\ && +331\,{\mathrm{e}}^{-4\,\beta \,h} \,{\mathrm{e}}^{4\,\beta \,\mu } +8\,{\mathrm{e}}^{4\,\beta \,h} \,{\mathrm{e}}^{4\,\beta \,\mu } +1078\,{\mathrm{e}}^{-6\,\beta \,h} \,{\mathrm{e}}^{2\,\beta \,\mu } +274\,{\mathrm{e}}^{-8\,\beta \,h} \,{\mathrm{e}}^{\beta \,\mu } +703\,{\mathrm{e}}^{-6\,\beta \,h} \,{\mathrm{e}}^{3\,\beta \,\mu } +7\,{\mathrm{e}}^{6\,\beta \,h} \,{\mathrm{e}}^{3\,\beta \,\mu } 
\\ && +189\,{\mathrm{e}}^{-6\,\beta \,h} \,{\mathrm{e}}^{4\,\beta \,\mu } +406\,{\mathrm{e}}^{-8\,\beta \,h} \,{\mathrm{e}}^{2\,\beta \,\mu } +82\,{\mathrm{e}}^{-10\,\beta \,h} \,{\mathrm{e}}^{\beta \,\mu } +266\,{\mathrm{e}}^{-8\,\beta \,h} \,{\mathrm{e}}^{3\,\beta \,\mu } +56\,{\mathrm{e}}^{-8\,\beta \,h} \,{\mathrm{e}}^{4\,\beta \,\mu } +88\,{\mathrm{e}}^{-10\,\beta \,h} \,{\mathrm{e}}^{2\,\beta \,\mu } 
\\ && +12\,{\mathrm{e}}^{-12\,\beta \,h} \,{\mathrm{e}}^{\beta \,\mu } +56\,{\mathrm{e}}^{-10\,\beta \,h} \,{\mathrm{e}}^{3\,\beta \,\mu } +9\,{\mathrm{e}}^{-10\,\beta \,h} \,{\mathrm{e}}^{4\,\beta \,\mu } +12\,{\mathrm{e}}^{-12\,\beta \,h} \,{\mathrm{e}}^{2\,\beta \,\mu } +4\,{\mathrm{e}}^{-12\,\beta \,h} \,{\mathrm{e}}^{3\,\beta \,\mu } +{\mathrm{e}}^{-12\,\beta \,h} \,{\mathrm{e}}^{4\,\beta \,\mu } 
\\ &&+{\mathrm{e}}^{-14\,\beta \,h} \,{\mathrm{e}}^{2\,\beta \,\mu }
\end{eqnarray*}

\begin{eqnarray*}
c \cdot &&e^{-4 \beta h'}  \overset{!}{=} e^{- \beta \hat{H'}(-1,-1,-1,-1)}
\\ = && 
1087\,{\mathrm{e}}^{-2\,\beta \,h} +199\,{\mathrm{e}}^{2\,\beta \,h} +1184\,{\mathrm{e}}^{-4\,\beta \,h} +26\,{\mathrm{e}}^{4\,\beta \,h} +891\,{\mathrm{e}}^{-6\,\beta \,h} +482\,{\mathrm{e}}^{-8\,\beta \,h} +191\,{\mathrm{e}}^{-10\,\beta \,h} +60\,{\mathrm{e}}^{-12\,\beta \,h} +19\,{\mathrm{e}}^{-14\,\beta \,h} 
\\ && +6\,{\mathrm{e}}^{-16\,\beta \,h} +{\mathrm{e}}^{-18\,\beta \,h} +52\,{\mathrm{e}}^{\beta \,\mu } +210\,{\mathrm{e}}^{2\,\beta \,\mu } +68\,{\mathrm{e}}^{3\,\beta \,\mu } +32\,{\mathrm{e}}^{4\,\beta \,\mu } +360\,{\mathrm{e}}^{-2\,\beta \,h} \,{\mathrm{e}}^{\beta \,\mu } +693\,{\mathrm{e}}^{-2\,\beta \,h} \,{\mathrm{e}}^{2\,\beta \,\mu } +27\,{\mathrm{e}}^{2\,\beta \,h} \,{\mathrm{e}}^{2\,\beta \,\mu } 
\\ && +948\,{\mathrm{e}}^{-4\,\beta \,h} \,{\mathrm{e}}^{\beta \,\mu } +444\,{\mathrm{e}}^{-2\,\beta \,h} \,{\mathrm{e}}^{3\,\beta \,\mu } +128\,{\mathrm{e}}^{-2\,\beta \,h} \,{\mathrm{e}}^{4\,\beta \,\mu } +4\,{\mathrm{e}}^{2\,\beta \,h} \,{\mathrm{e}}^{4\,\beta \,\mu } +1332\,{\mathrm{e}}^{-4\,\beta \,h} \,{\mathrm{e}}^{2\,\beta \,\mu } +1240\,{\mathrm{e}}^{-6\,\beta \,h} \,{\mathrm{e}}^{\beta \,\mu }
\\ && +1100\,{\mathrm{e}}^{-4\,\beta \,h} \,{\mathrm{e}}^{3\,\beta \,\mu } +294\,{\mathrm{e}}^{-4\,\beta \,h} \,{\mathrm{e}}^{4\,\beta \,\mu } +1724\,{\mathrm{e}}^{-6\,\beta \,h} \,{\mathrm{e}}^{2\,\beta \,\mu } +944\,{\mathrm{e}}^{-8\,\beta \,h} \,{\mathrm{e}}^{\beta \,\mu } +1360\,{\mathrm{e}}^{-6\,\beta \,h} \,{\mathrm{e}}^{3\,\beta \,\mu } +374\,{\mathrm{e}}^{-6\,\beta \,h} \,{\mathrm{e}}^{4\,\beta \,\mu }
\\ && +1516\,{\mathrm{e}}^{-8\,\beta \,h} \,{\mathrm{e}}^{2\,\beta \,\mu } +516\,{\mathrm{e}}^{-10\,\beta \,h} \,{\mathrm{e}}^{\beta \,\mu } +984\,{\mathrm{e}}^{-8\,\beta \,h} \,{\mathrm{e}}^{3\,\beta \,\mu } +260\,{\mathrm{e}}^{-8\,\beta \,h} \,{\mathrm{e}}^{4\,\beta \,\mu } +832\,{\mathrm{e}}^{-10\,\beta \,h} \,{\mathrm{e}}^{2\,\beta \,\mu } +224\,{\mathrm{e}}^{-12\,\beta \,h} \,{\mathrm{e}}^{\beta \,\mu } 
\\ && +456\,{\mathrm{e}}^{-10\,\beta \,h} \,{\mathrm{e}}^{3\,\beta \,\mu } +90\,{\mathrm{e}}^{-10\,\beta \,h} \,{\mathrm{e}}^{4\,\beta \,\mu } +268\,{\mathrm{e}}^{-12\,\beta \,h} \,{\mathrm{e}}^{2\,\beta \,\mu } +64\,{\mathrm{e}}^{-14\,\beta \,h} \,{\mathrm{e}}^{\beta \,\mu } +128\,{\mathrm{e}}^{-12\,\beta \,h} \,{\mathrm{e}}^{3\,\beta \,\mu } +12\,{\mathrm{e}}^{-12\,\beta \,h} \,{\mathrm{e}}^{4\,\beta \,\mu } 
\\ && +48\,{\mathrm{e}}^{-14\,\beta \,h} \,{\mathrm{e}}^{2\,\beta \,\mu } +8\,{\mathrm{e}}^{-16\,\beta \,h} \,{\mathrm{e}}^{\beta \,\mu } +16\,{\mathrm{e}}^{-14\,\beta \,h} \,{\mathrm{e}}^{3\,\beta \,\mu } +4\,{\mathrm{e}}^{-16\,\beta \,h} \,{\mathrm{e}}^{2\,\beta \,\mu } +630
\end{eqnarray*}

We drop all higher order terms and set $$\hat{H'} \overset{!}{=} - h' \sum_{\langle\mathbf{i}, \mathbf{j}\rangle}\hat{\tau}^x_{\langle\mathbf{i}, \mathbf{j}\rangle} 
- \mu' \sum_{\mathbf{j}} \hat{n}_{\mathbf{j}} + \text{constant} .$$ 
Solving for $\beta h'$ and $\beta \mu'$ gives
\begin{eqnarray}
\beta \mu'
= \frac{1}{2} \cdot 
\log(\frac{c e^{2\beta h'} e^{\beta \mu'} \cdot c e^{-2\beta h'} e^{\beta \mu'}}{c\cdot c})
= \frac{1}{2} \cdot 
\log(\frac{e^{- \beta \hat{H'}(-1,1,1,1)}\cdot e^{- \beta \hat{H'}(-1,-1,-1,1)}}{e^{- \beta \hat{H'}(-1,-1,1,1)}\cdot e^{- \beta \hat{H'}(-1,1,-1,1)}}) \label{eqn:mu_flow}
\end{eqnarray}
and
\begin{eqnarray} \beta h' 
= \frac{1}{8} \cdot \log(\frac{c \cdot e^{4 \beta h'}}{ c \cdot e^{ - 4 \beta h'}})
= \frac{1}{8} \cdot \log(\frac{e^{- \beta \hat{H'}(1,1,1,1)}}{ e^{- \beta \hat{H'}(-1,-1,-1,-1)}}). \label{eqn:h_flow}
\end{eqnarray}

This results in the flow diagram shown in Fig.~\ref{fig:RGflow_2D} and \ref{fig:RGflow_2D_wlims}. In particular, all $(\beta h, \beta \mu)$ with $\beta \mu > - \infty$ and $\beta h > 0$ flow to the non-percolating fixed point at $\beta \mu = 0$, $\beta h = \infty$.

We can analytically take limits of the flow equations to derive the behavior on each of the axes. 

In the limit $\beta h = 0$ we get
$$\beta h' =
\frac{1}{8} \cdot
\log \left(\frac{4068\,{\mathrm{e}}^{\beta \,\mu } +6750\,{\mathrm{e}}^{2\,\beta \,\mu } +4844\,{\mathrm{e}}^{3\,\beta \,\mu } +1290\,{\mathrm{e}}^{4\,\beta \,\mu } +4968}{4356\,{\mathrm{e}}^{\beta \,\mu } +6654\,{\mathrm{e}}^{2\,\beta \,\mu } +4556\,{\mathrm{e}}^{3\,\beta \,\mu } +1194\,{\mathrm{e}}^{4\,\beta \,\mu } +4776}\right)
$$
 and 
\begin{eqnarray*}
\beta \mu' = \frac{1}{2} \cdot 
\log \left( 
\frac{4\,{\mathrm{e}}^{2\,\beta \,\mu } \,{\left(2769\,{\mathrm{e}}^{\beta \,\mu } +1770\,{\mathrm{e}}^{2\,\beta \,\mu } +427\,{\mathrm{e}}^{3\,\beta \,\mu } +1966\right)}\,{\left(3369\,{\mathrm{e}}^{\beta \,\mu } +2290\,{\mathrm{e}}^{2\,\beta \,\mu } +579\,{\mathrm{e}}^{3\,\beta \,\mu } +2214\right)}}
{\splitfrac{\left(3956\,{\mathrm{e}}^{\beta \,\mu } +5550\,{\mathrm{e}}^{2\,\beta \,\mu } +3420\,{\mathrm{e}}^{3\,\beta \,\mu } +794\,{\mathrm{e}}^{4\,\beta \,\mu } +1128\right)}
{\phantom{4452\,{\mathrm{e}}^{\beta \,\mu } +6750\,{\mathrm{e}}^{2\,\beta \,\mu }} \cdot \left(4452\,{\mathrm{e}}^{\beta \,\mu } +6750\,{\mathrm{e}}^{2\,\beta \,\mu } +4460\,{\mathrm{e}}^{3\,\beta \,\mu } +1098\,{\mathrm{e}}^{4\,\beta \,\mu } +1160\right)}}
\right).
\end{eqnarray*}
Thus we (approximately) stay on this axis and flow towards the fixed point $(\beta h, \beta \mu)=(0, -0.077)$. The system percolates at the point $(\beta h, \beta \mu)=(0, 0)$ and flows from there to this fixed point, implying percolation for all $(\beta h, \beta \mu)=(0, \beta \mu)$.

In the limit $\beta \mu = -\infty$ (pure gauge) we get $\beta \mu ' = -\infty$ and 
\begin{eqnarray*}
\beta h' = \frac{1}{8} \cdot 
\log \left( 
\frac{
\splitfrac{
{\mathrm{e}}^{12\,\beta \,h} \,
(56\,{\mathrm{e}}^{2\,\beta \,h} +237\,{\mathrm{e}}^{4\,\beta \,h} +517\,{\mathrm{e}}^{6\,\beta \,h} +648\,{\mathrm{e}}^{8\,\beta \,h} +512\,{\mathrm{e}}^{10\,\beta \,h} +303\,{\mathrm{e}}^{12\,\beta \,h} 
}{+139\,{\mathrm{e}}^{14\,\beta \,h} +46\,{\mathrm{e}}^{16\,\beta \,h} +14\,{\mathrm{e}}^{18\,\beta \,h} +5\,{\mathrm{e}}^{20\,\beta \,h} +{\mathrm{e}}^{22\,\beta \,h} +6)}}
{\splitfrac{\quad(5\,{\mathrm{e}}^{2\,\beta \,h} +14\,{\mathrm{e}}^{4\,\beta \,h} +46\,{\mathrm{e}}^{6\,\beta \,h} +145\,{\mathrm{e}}^{8\,\beta \,h} +337\,{\mathrm{e}}^{10\,\beta \,h} +554\,{\mathrm{e}}^{12\,\beta \,h} 
}{+630\,{\mathrm{e}}^{14\,\beta \,h} +457\,{\mathrm{e}}^{16\,\beta \,h} +173\,{\mathrm{e}}^{18\,\beta \,h} +26\,{\mathrm{e}}^{20\,\beta \,h} +1)}}
\right).
\end{eqnarray*}
Thus for $\beta h > 0.29$ and for $\beta h < 0.074$ we have $\beta h ' > \beta h$ and for $0.074 < \beta h < 0.29$ we have $\beta h' < \beta h$.

In the limit $\beta \mu = 0$ (Bernoulli percolation), the flow equations give
\begin{eqnarray*}
\beta h' = \frac{1}{8} \cdot 
\log \left( 
\frac{
\splitfrac{
324\,{\mathrm{e}}^{-2\,\beta \,h} +2780\,{\mathrm{e}}^{2\,\beta \,h} +62\,{\mathrm{e}}^{-4\,\beta \,h} +4846\,{\mathrm{e}}^{4\,\beta \,h} +6\,{\mathrm{e}}^{-6\,\beta \,h} +5585\,{\mathrm{e}}^{6\,\beta \,h} 
}{+4234\,{\mathrm{e}}^{8\,\beta \,h} +2115\,{\mathrm{e}}^{10\,\beta \,h} +692\,{\mathrm{e}}^{12\,\beta \,h} +145\,{\mathrm{e}}^{14\,\beta \,h} +18\,{\mathrm{e}}^{16\,\beta \,h} +{\mathrm{e}}^{18\,\beta \,h} +1112
}}
{
\splitfrac{
2712\,{\mathrm{e}}^{-2\,\beta \,h} +230\,{\mathrm{e}}^{2\,\beta \,h} +4858\,{\mathrm{e}}^{-4\,\beta \,h} +26\,{\mathrm{e}}^{4\,\beta \,h} +5589\,{\mathrm{e}}^{-6\,\beta \,h} +4186\,{\mathrm{e}}^{-8\,\beta \,h} 
}{+2085\,{\mathrm{e}}^{-10\,\beta \,h} +692\,{\mathrm{e}}^{-12\,\beta \,h} +147\,{\mathrm{e}}^{-14\,\beta \,h} +18\,{\mathrm{e}}^{-16\,\beta \,h} +{\mathrm{e}}^{-18\,\beta \,h} +992}}
\right)
\end{eqnarray*}
and thus $\beta h' > \beta h $ for all $\beta h > 0$ on this axis.

For the limit $\beta h \to \infty$ there is only one allowed configuration - the configuration with zero electric strings. Thus the RG flow is not uniquely defined in this limit. Because the flow equations give $|\beta \mu '| < |\beta \mu|$ for large enough $\beta h$, we define the flow in this limit the same way, e.g. $\beta \mu' := \beta \mu +1$.

Altogether we obtain the RG flow depicted in Fig.~\ref{fig:RGflow_2D_wlims}b and \ref{fig:RGflow_2D}.

\section{Correspondence to site-bond percolation} \label{sec:RG_appendix_sitebondperc}

We give a rough bound of the consequences which percolation corrections have on the confinement phase diagram by comparing to a site-bond percolation problem.

Note that $|\delta \rho| \lesssim 0.08$ for all $(\beta h, \beta \mu)$ with the maximum value being attained at $(\beta h, \beta \mu) \to (0,-\infty)$. In the uncorrelated limit, the effect on the confinement phase diagram corresponds to a site-bond percolation problem. The site-bond percolation problem considers a lattice on which both sites and bonds can be randomly occupied \cite{Frisch1963}. The correspondence is then established by assigning occupancy to macro-sites in a given RG step according to whether the percolation carries through the corresponding micro-configuration. Specifically, we say for a bond-percolating block configuration that the central site is occupied if and only if the micro-configuration percolates. (Note that the occupation of sites defined here is distinct from the charges defined above.) In the worst case we thus have a site occupation probability of $1- \delta \rho \approx 0.92$. On the curve of critical Bernoulli site-bond percolation this corresponds to a bond occupation probability of $0.56$ \cite{Yanuka1990,Tarasevich1999}.

It follows that the critical bond occupation probability is shifted in the uncorrelated system by at most $0.06$. Assuming that the effect is similar in the correlated system, we can conclude that the percolation corrections do not affect the RG flow for large enough $\beta h$ (small $T/h$).



\section{Details of percolation flow}

The percolation change is defined by the difference in percolation probabilities of the the renormalized configuration and the original configuration. To calculate this, we check for each micro-configuration on a block whether or not it percolates and compare this to the percolation of the assigned macro-configuration. 

\begin{figure}
\centering
\includegraphics[width = 0.6\textwidth]{\empty 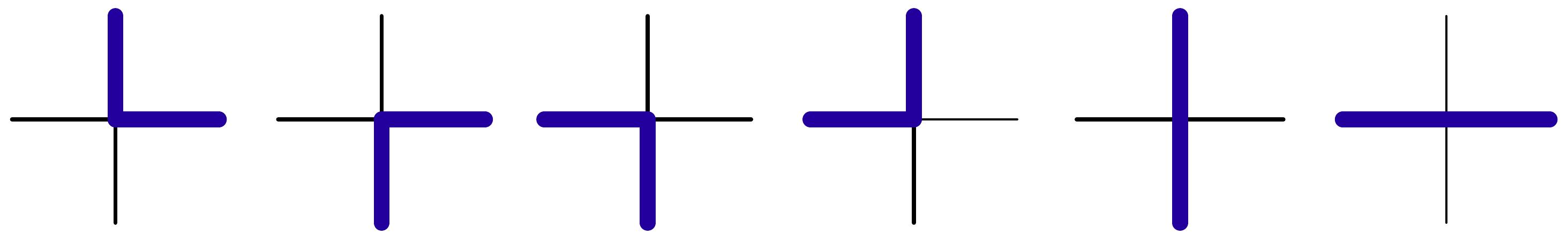}
\caption{\label{fig:PercPaths} \textbf{Percolating paths through a block.}
There are several percolating paths through a block with respect to which the percolation change can be defined. In our calculations, we take the average over all such paths.
}
\end{figure}

The percolation flow depends on the path with respect to which we define percolation, e.g. from left to right or from the bottom to the right. To simplify, we take the average over all such paths, shown in Fig.~\ref{fig:PercPaths}.

The resulting percolation change is
\begin{eqnarray*}
\delta \rho(\beta h, \beta \mu)
= && \frac{1}{Z(\beta h, \beta \mu)} \cdot \frac{1}{6}\Big(
2470\,{\mathrm{e}}^{-2\,\beta \,h} -68\,{\mathrm{e}}^{2\,\beta \,h} +2630\,{\mathrm{e}}^{-4\,\beta \,h} -418\,{\mathrm{e}}^{4\,\beta \,h} +1712\,{\mathrm{e}}^{-6\,\beta \,h} -277\,{\mathrm{e}}^{6\,\beta \,h} +682\,{\mathrm{e}}^{-8\,\beta \,h} 
\\ && -124\,{\mathrm{e}}^{8\,\beta \,h} +170\,{\mathrm{e}}^{-10\,\beta \,h} -44\,{\mathrm{e}}^{10\,\beta \,h} +36\,{\mathrm{e}}^{-12\,\beta \,h} -10\,{\mathrm{e}}^{12\,\beta \,h} +6\,{\mathrm{e}}^{-14\,\beta \,h} -{\mathrm{e}}^{14\,\beta \,h} -1712\,{\mathrm{e}}^{\beta \,\mu } -1108\,{\mathrm{e}}^{2\,\beta \,\mu } 
\\ && -126\,{\mathrm{e}}^{3\,\beta \,\mu } +220\,{\mathrm{e}}^{4\,\beta \,\mu } -1360\,{\mathrm{e}}^{-2\,\beta \,h} \,{\mathrm{e}}^{\beta \,\mu } -1646\,{\mathrm{e}}^{2\,\beta \,h} \,{\mathrm{e}}^{\beta \,\mu } -977\,{\mathrm{e}}^{-2\,\beta \,h} \,{\mathrm{e}}^{2\,\beta \,\mu } -1517\,{\mathrm{e}}^{2\,\beta \,h} \,{\mathrm{e}}^{2\,\beta \,\mu } 
\\ && -808\,{\mathrm{e}}^{-4\,\beta \,h} \,{\mathrm{e}}^{\beta \,\mu } -1290\,{\mathrm{e}}^{4\,\beta \,h} \,{\mathrm{e}}^{\beta \,\mu } -170\,{\mathrm{e}}^{-2\,\beta \,h} \,{\mathrm{e}}^{3\,\beta \,\mu } -208\,{\mathrm{e}}^{2\,\beta \,h} \,{\mathrm{e}}^{3\,\beta \,\mu } +120\,{\mathrm{e}}^{-2\,\beta \,h} \,{\mathrm{e}}^{4\,\beta \,\mu } +156\,{\mathrm{e}}^{2\,\beta \,h} \,{\mathrm{e}}^{4\,\beta \,\mu } 
\\ && -1172\,{\mathrm{e}}^{-4\,\beta \,h} \,{\mathrm{e}}^{2\,\beta \,\mu } -1316\,{\mathrm{e}}^{4\,\beta \,h} \,{\mathrm{e}}^{2\,\beta \,\mu } -396\,{\mathrm{e}}^{-6\,\beta \,h} \,{\mathrm{e}}^{\beta \,\mu } -786\,{\mathrm{e}}^{6\,\beta \,h} \,{\mathrm{e}}^{\beta \,\mu } -400\,{\mathrm{e}}^{-4\,\beta \,h} \,{\mathrm{e}}^{3\,\beta \,\mu } -260\,{\mathrm{e}}^{4\,\beta \,h} \,{\mathrm{e}}^{3\,\beta \,\mu }
\\ && -64\,{\mathrm{e}}^{-4\,\beta \,h} \,{\mathrm{e}}^{4\,\beta \,\mu } +26\,{\mathrm{e}}^{4\,\beta \,h} \,{\mathrm{e}}^{4\,\beta \,\mu } -933\,{\mathrm{e}}^{-6\,\beta \,h} \,{\mathrm{e}}^{2\,\beta \,\mu } -645\,{\mathrm{e}}^{6\,\beta \,h} \,{\mathrm{e}}^{2\,\beta \,\mu } -138\,{\mathrm{e}}^{-8\,\beta \,h} \,{\mathrm{e}}^{\beta \,\mu } -318\,{\mathrm{e}}^{8\,\beta \,h} \,{\mathrm{e}}^{\beta \,\mu } 
\\ && -580\,{\mathrm{e}}^{-6\,\beta \,h} \,{\mathrm{e}}^{3\,\beta \,\mu } -194\,{\mathrm{e}}^{6\,\beta \,h} \,{\mathrm{e}}^{3\,\beta \,\mu } -170\,{\mathrm{e}}^{-6\,\beta \,h} \,{\mathrm{e}}^{4\,\beta \,\mu } -26\,{\mathrm{e}}^{6\,\beta \,h} \,{\mathrm{e}}^{4\,\beta \,\mu } -400\,{\mathrm{e}}^{-8\,\beta \,h} \,{\mathrm{e}}^{2\,\beta \,\mu } -184\,{\mathrm{e}}^{8\,\beta \,h} \,{\mathrm{e}}^{2\,\beta \,\mu } 
\\ && -22\,{\mathrm{e}}^{-10\,\beta \,h} \,{\mathrm{e}}^{\beta \,\mu } -70\,{\mathrm{e}}^{10\,\beta \,h} \,{\mathrm{e}}^{\beta \,\mu } -470\,{\mathrm{e}}^{-8\,\beta \,h} \,{\mathrm{e}}^{3\,\beta \,\mu } -76\,{\mathrm{e}}^{8\,\beta \,h} \,{\mathrm{e}}^{3\,\beta \,\mu } -132\,{\mathrm{e}}^{-8\,\beta \,h} \,{\mathrm{e}}^{4\,\beta \,\mu } -14\,{\mathrm{e}}^{8\,\beta \,h} \,{\mathrm{e}}^{4\,\beta \,\mu } 
\\ && -86\,{\mathrm{e}}^{-10\,\beta \,h} \,{\mathrm{e}}^{2\,\beta \,\mu } -34\,{\mathrm{e}}^{10\,\beta \,h} \,{\mathrm{e}}^{2\,\beta \,\mu } -6\,{\mathrm{e}}^{12\,\beta \,h} \,{\mathrm{e}}^{\beta \,\mu } -188\,{\mathrm{e}}^{-10\,\beta \,h} \,{\mathrm{e}}^{3\,\beta \,\mu } -12\,{\mathrm{e}}^{10\,\beta \,h} \,{\mathrm{e}}^{3\,\beta \,\mu } -66\,{\mathrm{e}}^{-10\,\beta \,h} \,{\mathrm{e}}^{4\,\beta \,\mu } 
\\ && -2\,{\mathrm{e}}^{10\,\beta \,h} \,{\mathrm{e}}^{4\,\beta \,\mu } 
-8\,{\mathrm{e}}^{-12\,\beta \,h} \,{\mathrm{e}}^{2\,\beta \,\mu } -4\,{\mathrm{e}}^{12\,\beta \,h} \,{\mathrm{e}}^{2\,\beta \,\mu } -28\,{\mathrm{e}}^{-12\,\beta \,h} \,{\mathrm{e}}^{3\,\beta \,\mu } -24\,{\mathrm{e}}^{-12\,\beta \,h} \,{\mathrm{e}}^{4\,\beta \,\mu } -4\,{\mathrm{e}}^{-14\,\beta \,h} \,{\mathrm{e}}^{4\,\beta \,\mu } 
\\ && +1168
\Big)
\end{eqnarray*}
and is normalized by the partition sum
\begin{eqnarray*}
Z(\beta h, \beta \mu)
= &&~
2796\,{\mathrm{e}}^{-2\,\beta \,h} +2796\,{\mathrm{e}}^{2\,\beta \,h} +1924\,{\mathrm{e}}^{-4\,\beta \,h} +1924\,{\mathrm{e}}^{4\,\beta \,h} +1086\,{\mathrm{e}}^{-6\,\beta \,h} +1086\,{\mathrm{e}}^{6\,\beta \,h} +512\,{\mathrm{e}}^{-8\,\beta \,h} +512\,{\mathrm{e}}^{8\,\beta \,h} 
\\ && +194\,{\mathrm{e}}^{-10\,\beta \,h} 
+194\,{\mathrm{e}}^{10\,\beta \,h} +60\,{\mathrm{e}}^{-12\,\beta \,h} +60\,{\mathrm{e}}^{12\,\beta \,h} +19\,{\mathrm{e}}^{-14\,\beta \,h} +19\,{\mathrm{e}}^{14\,\beta \,h} +6\,{\mathrm{e}}^{-16\,\beta \,h} +6\,{\mathrm{e}}^{16\,\beta \,h} 
\\ && +{\mathrm{e}}^{-18\,\beta \,h} +{\mathrm{e}}^{18\,\beta \,h} +12336\,{\mathrm{e}}^{\beta \,\mu } +18360\,{\mathrm{e}}^{2\,\beta \,\mu } +11904\,{\mathrm{e}}^{3\,\beta \,\mu } +2832\,{\mathrm{e}}^{4\,\beta \,\mu } +11072\,{\mathrm{e}}^{-2\,\beta \,h} \,{\mathrm{e}}^{\beta \,\mu } 
\\ && +11072\,{\mathrm{e}}^{2\,\beta \,h} \,{\mathrm{e}}^{\beta \,\mu } +16444\,{\mathrm{e}}^{-2\,\beta \,h} \,{\mathrm{e}}^{2\,\beta \,\mu } +16444\,{\mathrm{e}}^{2\,\beta \,h} \,{\mathrm{e}}^{2\,\beta \,\mu } +7936\,{\mathrm{e}}^{-4\,\beta \,h} \,{\mathrm{e}}^{\beta \,\mu } +7936\,{\mathrm{e}}^{4\,\beta \,h} \,{\mathrm{e}}^{\beta \,\mu } 
\\ && +10800\,{\mathrm{e}}^{-2\,\beta \,h} \,{\mathrm{e}}^{3\,\beta \,\mu } +10800\,{\mathrm{e}}^{2\,\beta \,h} \,{\mathrm{e}}^{3\,\beta \,\mu } +2646\,{\mathrm{e}}^{-2\,\beta \,h} \,{\mathrm{e}}^{4\,\beta \,\mu } +2646\,{\mathrm{e}}^{2\,\beta \,h} \,{\mathrm{e}}^{4\,\beta \,\mu } +11848\,{\mathrm{e}}^{-4\,\beta \,h} \,{\mathrm{e}}^{2\,\beta \,\mu } 
\\ && +11848\,{\mathrm{e}}^{4\,\beta \,h} \,{\mathrm{e}}^{2\,\beta \,\mu } +4480\,{\mathrm{e}}^{-6\,\beta \,h} \,{\mathrm{e}}^{\beta \,\mu } +4480\,{\mathrm{e}}^{6\,\beta \,h} \,{\mathrm{e}}^{\beta \,\mu } +8032\,{\mathrm{e}}^{-4\,\beta \,h} \,{\mathrm{e}}^{3\,\beta \,\mu } +8032\,{\mathrm{e}}^{4\,\beta \,h} \,{\mathrm{e}}^{3\,\beta \,\mu } 
\\ && +2084\,{\mathrm{e}}^{-4\,\beta \,h} \,{\mathrm{e}}^{4\,\beta \,\mu } +2084\,{\mathrm{e}}^{4\,\beta \,h} \,{\mathrm{e}}^{4\,\beta \,\mu } +6900\,{\mathrm{e}}^{-6\,\beta \,h} \,{\mathrm{e}}^{2\,\beta \,\mu } +6900\,{\mathrm{e}}^{6\,\beta \,h} \,{\mathrm{e}}^{2\,\beta \,\mu } +2016\,{\mathrm{e}}^{-8\,\beta \,h} \,{\mathrm{e}}^{\beta \,\mu } 
\\ && +2016\,{\mathrm{e}}^{8\,\beta \,h} \,{\mathrm{e}}^{\beta \,\mu } +4816\,{\mathrm{e}}^{-6\,\beta \,h} \,{\mathrm{e}}^{3\,\beta \,\mu } +4816\,{\mathrm{e}}^{6\,\beta \,h} \,{\mathrm{e}}^{3\,\beta \,\mu } +1282\,{\mathrm{e}}^{-6\,\beta \,h} \,{\mathrm{e}}^{4\,\beta \,\mu } +1282\,{\mathrm{e}}^{6\,\beta \,h} \,{\mathrm{e}}^{4\,\beta \,\mu } 
\\ && +3232\,{\mathrm{e}}^{-8\,\beta \,h} \,{\mathrm{e}}^{2\,\beta \,\mu } +3232\,{\mathrm{e}}^{8\,\beta \,h} \,{\mathrm{e}}^{2\,\beta \,\mu } +768\,{\mathrm{e}}^{-10\,\beta \,h} \,{\mathrm{e}}^{\beta \,\mu } +768\,{\mathrm{e}}^{10\,\beta \,h} \,{\mathrm{e}}^{\beta \,\mu } +2240\,{\mathrm{e}}^{-8\,\beta \,h} \,{\mathrm{e}}^{3\,\beta \,\mu } 
\\ && +2240\,{\mathrm{e}}^{8\,\beta \,h} \,{\mathrm{e}}^{3\,\beta \,\mu } +568\,{\mathrm{e}}^{-8\,\beta \,h} \,{\mathrm{e}}^{4\,\beta \,\mu } +568\,{\mathrm{e}}^{8\,\beta \,h} \,{\mathrm{e}}^{4\,\beta \,\mu } +1180\,{\mathrm{e}}^{-10\,\beta \,h} \,{\mathrm{e}}^{2\,\beta \,\mu } +1180\,{\mathrm{e}}^{10\,\beta \,h} \,{\mathrm{e}}^{2\,\beta \,\mu } 
\\ && +256\,{\mathrm{e}}^{-12\,\beta \,h} \,{\mathrm{e}}^{\beta \,\mu } +256\,{\mathrm{e}}^{12\,\beta \,h} \,{\mathrm{e}}^{\beta \,\mu } +752\,{\mathrm{e}}^{-10\,\beta \,h} \,{\mathrm{e}}^{3\,\beta \,\mu } +752\,{\mathrm{e}}^{10\,\beta \,h} \,{\mathrm{e}}^{3\,\beta \,\mu } +166\,{\mathrm{e}}^{-10\,\beta \,h} \,{\mathrm{e}}^{4\,\beta \,\mu } 
\\ && +166\,{\mathrm{e}}^{10\,\beta \,h} \,{\mathrm{e}}^{4\,\beta \,\mu } +312\,{\mathrm{e}}^{-12\,\beta \,h} \,{\mathrm{e}}^{2\,\beta \,\mu } +312\,{\mathrm{e}}^{12\,\beta \,h} \,{\mathrm{e}}^{2\,\beta \,\mu } +64\,{\mathrm{e}}^{-14\,\beta \,h} \,{\mathrm{e}}^{\beta \,\mu } +64\,{\mathrm{e}}^{14\,\beta \,h} \,{\mathrm{e}}^{\beta \,\mu } +160\,{\mathrm{e}}^{-12\,\beta \,h} \,{\mathrm{e}}^{3\,\beta \,\mu } 
\\ && +160\,{\mathrm{e}}^{12\,\beta \,h} \,{\mathrm{e}}^{3\,\beta \,\mu } +28\,{\mathrm{e}}^{-12\,\beta \,h} \,{\mathrm{e}}^{4\,\beta \,\mu } +28\,{\mathrm{e}}^{12\,\beta \,h} \,{\mathrm{e}}^{4\,\beta \,\mu } +52\,{\mathrm{e}}^{-14\,\beta \,h} \,{\mathrm{e}}^{2\,\beta \,\mu } +52\,{\mathrm{e}}^{14\,\beta \,h} \,{\mathrm{e}}^{2\,\beta \,\mu } +8\,{\mathrm{e}}^{-16\,\beta \,h} \,{\mathrm{e}}^{\beta \,\mu } 
\\ && +8\,{\mathrm{e}}^{16\,\beta \,h} \,{\mathrm{e}}^{\beta \,\mu } +16\,{\mathrm{e}}^{-14\,\beta \,h} \,{\mathrm{e}}^{3\,\beta \,\mu } +16\,{\mathrm{e}}^{14\,\beta \,h} \,{\mathrm{e}}^{3\,\beta \,\mu } +2\,{\mathrm{e}}^{-14\,\beta \,h} \,{\mathrm{e}}^{4\,\beta \,\mu } +2\,{\mathrm{e}}^{14\,\beta \,h} \,{\mathrm{e}}^{4\,\beta \,\mu } +4\,{\mathrm{e}}^{-16\,\beta \,h} \,{\mathrm{e}}^{2\,\beta \,\mu } 
\\ && +4\,{\mathrm{e}}^{16\,\beta \,h} \,{\mathrm{e}}^{2\,\beta \,\mu } +3188
.
\end{eqnarray*}

To estimate the total percolation change $\Delta \rho (\beta h, \beta \mu) = \sum_{n= 0}^\infty \delta\rho_n$ for the flow starting at some point in parameter space, we sum over the percolation change from the RG steps starting at this point.

We note that $\delta \rho$ is positive only in the region with $\beta \mu \ll 0$ and $\beta h$ small enough. Because the flow from any point $(\beta h, \beta \mu) \neq (\beta h, -\infty)$ eventually leaves this region, $\Delta \rho$ cannot be positive for any such parameter values. For $\beta \mu = - \infty$ we find the percolation transition point ($\Delta \rho (\beta h, -\infty) = 0$) to be the largest $\beta h$ from which the flow ends in the region with positive $\delta \rho$. The transition thus occurs at $(\beta h, \beta \mu) = (1/3.48...,-\infty)$.  The exact critical point is at $(\beta h, \beta \mu) = (1/2.27...,-\infty)$.

\section{Data analysis with Monte-Carlo snapshots}
\label{sec:data_analysis}

In this section, we consider another method to analyze the renormalization group flow. 

We first perform Monte-Carlo simulations of the model (\ref{eqn:gcHam}). The RG steps are then applied directly on the resulting snapshots which yields a data set of smaller, renormalized snapshots. 
For coupling strengths $\beta h$, $\beta \mu$ in the vicinity of a fixed point, there should be little change in the snapshots generated this way when compared to Monte-Carlo snapshots with no RG step applied. For coupling parameters far from any fixed point, the RG step significantly changes the lattice configurations. This effect is more prominent after several RG steps are applied.

In order to define an objective metric of similarity in sets of lattice snapshots, we train a neural network to distinguish these datasets. The change from applying an RG step is then quantified by the success rate of this neural network in distinguishing the datasets.

Table \ref{table:snapshots} shows the results for a set of $10000$ Monte-Carlo snapshots with system size $75\cross 75$. The renormalization procedure from section (\ref{sec:RG}) asymptotically halves the system length at each step, but the block covering has constant losses at the edges. Thus, applying an RG step to a state of size $(4 k + 1)\cross(4 k + 1)$ yields a configuration with size $(2k -1)\cross(2k -1)$.

{\renewcommand{\arraystretch}{1.5}
\begin{table*} 
\begin{ruledtabular}
\begin{tabular}{c c c c}
percolating phase & non-percolating phase  & near RG fixed point & near exact fixed point \\ 
$T/ h =  +\infty$, $ T/ \mu = -0.5$ 
& $T/ h =  1.0 $, $ T/ \mu = -0.5$
& $T/ h =  3.48$, $ T/ \mu =  - 3.48 \cdot 10^{-4}$
& $T/ h =  2.27$, $ T/ \mu =  - 2.27 \cdot 10^{-4}$ \\
\hline 
  $80.45 \%$
& $49.45 \%$
& $72.50 \%$
& $77.10 \% $
\\
\end{tabular}
\end{ruledtabular}
\caption{\label{table:snapshots}\textbf{Distinction accuracies of the neural network.} 
We generate $10000$ Monte-Carlo snapshots with system size $75\cross 75$ and open boundary conditions. To combat boundary effects, a square of size $45\cross 45$ from the center of these snapshots is used. We apply two RG steps to the first $5000$ snapshots. The resulting $9\cross 9$ lattice configurations are compared to same size cutouts from applying one RG step to the remaining $5000$ snapshots. A neural network with three hidden layers is trained on $80\%$ of the data and tested on the remaining $20 \%$. We show the distinction accuracies of the trained neural network between the datasets with two RG steps and with one RG step.
}
\end{table*}}

We can see that the effect of the RG step is indiscernible for the non-percolating configurations, i.e. the neural network accuracy is no better than a random guess. At the chosen parameters, most of the configurations are empty and so the renormalization has little effect. We recognise this as the trivial non-percolating fixed point at $T/h = 0$.
In the percolating phase at $T/h = + \infty$, the high accuracy shows that a single RG step strongly affects the system, indicating that the flow hasn't yet reached a fixed point. 
The accuracies near the critical fixed point at $T/h = 3.48$, $T/\mu = 0$ are harder to interpret and we would have expected lower accuracies in this case. It is possible that the results are due to the smaller system size, effects from the open boundaries, or $T/\mu$ being nonzero. 

Overall this method yields no striking results for the model at hand. It is however interesting since it could easily be applied to analyze other models where snapshots are available e.g. from quantum Monte-Carlo simulations or experimental data.

\end{document}